\documentclass[fleqn,usenatbib]{mnras}

\usepackage{newtxtext,newtxmath}

\usepackage[T1]{fontenc}

\DeclareRobustCommand{\VAN}[3]{#2}
\let\VANthebibliography\thebibliography
\def\thebibliography{\DeclareRobustCommand{\VAN}[3]{##3}\VANthebibliography}

\usepackage{graphicx}	
\usepackage{amsmath}	
\usepackage{amssymb}	
\usepackage{wasysym}    
\usepackage{mathtools}  
\usepackage{caption}
\usepackage{subcaption}
\usepackage{soul}

\newcommand{\solm}{M$_{\astrosun}$}
\newcommand{\isolm}{M$_{\astrosun}^{-1}$}
\newcommand{\lya}{Ly\,$\alpha$}
\newcommand{\rast}{\color{red}\,$^*$}
\newcommand{\ke}{E_{\rm K}}
\newcommand{\kemin}{E_{\rm K, min}}
\newcommand{\kemax}{E_{\rm K, max}}
\newcommand{\kezero}{E_{\rm K, 0}}

\title[Cosmic Ray Heating and the 21-cm Signal]{Signatures of Cosmic Ray Heating in 21-cm Observables}

\author[T. Gessey-Jones et al.]{T. Gessey-Jones,$^{1,2}$\thanks{E-mail: tg400@cam.ac.uk}
A. Fialkov,$^{2,3}$
E. de Lera Acedo,$^{1,2}$
W. J. Handley,$^{1,2}$
and R. Barkana$^{4,5,6}$
\\
$^{1}$Astrophysics Group, Cavendish Laboratory, J. J. Thomson Avenue, Cambridge, CB3 0HE, UK\\
$^{2}$Kavli Institute for Cosmology, Madingley Road, Cambridge, CB3 0HA, UK\\
$^{3}$Institute of Astronomy, University of Cambridge, Madingley Road, Cambridge, CB3 0HA, UK\\
$^{4}$School of Physics and Astronomy, Tel-Aviv University, Tel-Aviv 69978, Israel \\ 
$^{5}$Institute for Advanced Study, 1 Einstein Drive, Princeton, New Jersey 08540, USA \\
$^{6}$Department of Astronomy and Astrophysics, University of California, Santa Cruz, California 95064, USA
}

\date{Accepted XXX. Received YYY; in original form ZZZ}

\pubyear{2023}

\begin{document}
\label{firstpage}
\pagerange{\pageref{firstpage}--\pageref{lastpage}}
\maketitle

\begin{abstract}
Cosmic rays generated by supernovae carry away a significant portion of the lifetime energy emission of their parent star, making them a plausible mechanism for heating the early universe intergalactic medium (IGM).
Following a review of the existing literature on cosmic ray heating, we develop a flexible model of this heating mechanism for use in 3D semi-numerical 21-cm signal simulations and conduct the first investigations of the signatures it imprints on the 21-cm power spectrum and tomographic maps. We find that cosmic ray heating of the IGM is short-ranged, leading to heating clustered around star-forming sites, and a sharp contrast between heated regions of 21-cm emission and unheated regions of absorption. 
This contrast results in greater small-scale power for cosmic ray heated scenarios compared to what is found for X-ray heating, thus suggesting a way to test the nature of IGM heating with future 21-cm observations. Finally, we find an unexpectedly rich thermal history in models where cosmic rays can only escape efficiently from low-mass halos, such as in scenarios where these energetic particles originate from population III star supernovae remnants. 
The interplay of heating and the Lyman-Werner feedback in these models can produce a local peak in the IGM kinetic temperature and, for a limited parameter range, a flattened absorption trough in the global 21-cm signal. 
\end{abstract}

\begin{keywords}
cosmic rays -- early Universe -- dark ages, reionization, first stars -- cosmology: theory
\end{keywords}



\section{Introduction}

Observations of supernova remnants (SNRs) by the \textit{Fermi} satellite~\citep{FERMI_2013} strongly support the theory that supernovae are a source of cosmic rays. These high-energy charged particles are estimated to carry away 10 to 50 per cent of the kinetic energy of the supernova shock~\citep{Drury_1989, Berezinskii_1990, Caprioli_2014, Sazonov_2015} and so represent a substantial portion of the lifetime energy output of massive stars. 

Due to the large amount of energy imparted into cosmic rays and the important role played by cosmic rays in determining the thermal and ionization state of the interstellar medium \citep[ISM, ][]{Schlickeiser_2002}, previous authors~\citep{Ginzburg_1966, Nath_1993, Sazonov_2015, Leite_2017, Jana_2018, Jana_2019, Bera_2022} have considered them as a potential mechanism for heating and ionizing the neutral intergalactic medium (IGM) at early times. For example,  \citet{Sazonov_2015} and \citet{Leite_2017} demonstrate that cosmic rays produced by Population III (Pop III) and Population II (Pop II) star supernovae may be able to heat the IGM above the cosmic microwave background (CMB) temperature by redshift  $15$ and $10$ respectively. However, neither study found cosmic rays contribute more than a few per cent to the ionization of the IGM.

Any efficient heating of the IGM before reionization is of particular interest in the context of the burgeoning field of 21-cm cosmology~\citep{Madau_1997, Furlanetto_2006, Pritchard_2012, Barkana_2016, Mesinger_2019}. 21-cm cosmology aims to probe the dark age of the Universe, cosmic dawn, and the epoch of reionization via the degree of emission or absorption at the 21-cm spectral line of neutral hydrogen. The strength of this emission or absorption traces the evolution of the thermal and ionization states of the IGM providing a way to test the astrophysical properties of the first stars and galaxies~\citep{Yajima_2015, Cohen_2016, Mirocha_2018, Mebane_2018, Tanaka_2018,  Schauer_2019, Mirocha_2019, Mebane_2020, Tanaka_2021, Munoz_2022, Gessey-Jones_2022}, and potentially the nature of dark matter~\citep{Barkana_2018, Munoz_2018, Fraser_2018, Fialkov_2018, Liu_2019, Munoz_2020, Jones_2021, Hibbard_2022, Barkana_2022}. Optimistically, cosmic ray heating may therefore provide another mechanism by which early stars and galaxies impact the surrounding IGM, allowing additional information to be extracted about their properties from the 21-cm signal. Pessimistically, the impacts of cosmic ray heating on the 21-cm signal may be degenerate with those of other physical processes of interest thereby weakening possible constraints from a 21-cm signal detection. We leave investigations of such degeneracies to a future study, as this paper is focused on the detailed modelling of this heating contribution and providing some illustrative examples. 

An additional motivation to explore the potential contribution of cosmic rays to the IGM heating is provided by the disputed global signal detected by the EDGES collaboration~\citep{EDGES, Hills_2018, SARAS3}. It was shown by \citet{Jana_2019} that cosmic ray heating prevents the radio background from supernovae alone being an explanation for the anomalous depth of the reported signal, while \citet{Bera_2022} argued models with cosmic ray heating alongside dark matter-baryon interactions could provide a potential explanation for some features of the best-fit absorption trough. It is, thus, evident that an understanding of the signatures of cosmic ray heating in the 21-cm signal may be necessary for the correct interpretation of the signal. 

The aforementioned studies into cosmic ray heating found it to be dominated by sub-relativistic cosmic ray protons. Due to these particles moving at less than the speed of light, and potentially diffusing throughout the IGM by scattering from magnetic irregularities, heating from cosmic rays is anticipated to be more clustered around star-forming sources than alternatives such as X-ray heating~\citep{Fialkov_2014b, Pacucci_2014}. \citet{Leite_2017} and \citet{Yokoyama_2023} both briefly discussed that this clustering of heating could result in specific features in the 21-cm power spectrum that would not be present in the case of an IGM heated by X-rays, potentially allowing a cosmic ray heated IGM to be distinguished from an X-ray heated one; however, no further study of these features was conducted. 

In this paper, we fill this gap in the literature by performing the first 21-cm signal simulations to model the spatial distribution of cosmic ray heating. 
This is achieved by extending an existing semi-numerical 21-cm signal simulation~\citep[e.g.,][]{Visbal_2012, Fialkov_2014, Reis_2020} with a parameterized model of cosmic ray heating. 
By using a semi-numerical simulation, rather than a globally averaged semi-analytic model like previous studies, we are able to simulate full 21-cm tomographic maps and thus calculate the 21-cm power spectrum.
Hence, allowing for the first quantitative discussions on the impact of cosmic ray heating on these observables, and on whether or not cosmic ray heating may be distinguishable from other heating mechanisms of the early IGM.
Furthermore, as the 21-cm signal is not linearly dependent on the IGM gas temperature, spatially resolved simulations may be necessary to produce accurate predictions of the 21-cm global signal, especially given the anticipated strongly clustered nature of cosmic ray heating. 
We are thus also able to assess the extent to which the globally uniform heating assumption has biased the conclusions reached by previous studies.
Our approach additionally enables us to investigate the sensitivity of cosmic ray heating to other uncertain aspects of the high redshift universe such as the star formation history and the strength of the Lyman-Werner (LW) feedback~\citep{Haiman_2000}. 
We hence also consider how the imprints of cosmic ray heating vary with the uncertain astrophysical properties of the first stars and galaxies, and thus verify the robustness of our conclusions.

We begin this paper by reviewing the modelling of cosmic ray heating in previous studies in section~\ref{sec:cr_heating}, which a reader familiar with the literature may want to skip in the interest of time. In section~\ref{sec:21cm} we then recap the theory of the 21-cm signal, outline our semi-numerical simulation code, and describe our parameterized model of cosmic ray heating. Using our extended simulations in section~\ref{sec:results} we detail our findings regarding the signatures of cosmic ray heating in various 21-cm observables and highlight unique features seen for specific types of cosmic ray heating scenarios. Finally, in section~\ref{sec:conclusions} we conclude with a summary of our results and discuss how they could be more generally relevant to the interpretation of the 21-cm signal.

\section{Cosmic Ray Heating of the IGM}
\label{sec:cr_heating}

A model of high-redshift cosmic ray heating requires several key ingredients: a production mechanism of cosmic rays in the early universe (subsection~\ref{ssec:cr_production}); a prescription for how these cosmic rays reach the IGM (subsection~\ref{ssec:cr_escape}); channels through which cosmic rays lose their energy (subsection~\ref{ssec:cr_energy_loss}); a way to convert this energy loss into the IGM heat (subsection~\ref{ssec:cr_heating}); and finally the heating distribution caused by the propagation of cosmic rays through the IGM (subsection~\ref{ssec:cr_propogation}). 

Approaches taken to each one of these modelling steps have differed between previous studies~\citep{Nath_1993, Stacy_2007, Sazonov_2015, Leite_2017, Jana_2018, Samui_2018, Jana_2019, Bera_2022}, due in part to the uncertainties surrounding cosmic ray astrophysics in the early universe. Here we briefly review the prior approaches taken when investigating the impacts of cosmic ray heating on the 21-cm signal.

\subsection{Production of cosmic rays} \label{ssec:cr_production}

As a dying massive star undergoes a supernova it injects $10^{51}$ to $10^{53}$\,erg of kinetic energy into ejected material~\citep{Woosley_1986, Heger_2002, Woosley_2007, Heger_2010, Woosley_2010, Whalen_2013b, Whalen_2013a, Chen_2014, Chen_2017}. The exact amount of energy is dependent on the type of supernovae and hence the mass of the star, with core-collapse supernovae yielding the lower end of the stated energy range and pair-instability supernovae producing the highest amount of energy. The expelled high-velocity material travels outwards from the star, eventually reaching the boundary of the photoevaporated region surrounding the star. As the high-velocity material meets this denser medium a shock forms. This shock then continues to expand outwards as a supernovae remnant (SNR), gradually losing its energy until it finally dissipates. 

A charged particle in the SNR can diffuse back and forth across the shock gaining energy each time, a process referred to as diffusive shock acceleration \citep[DSA,][]{Bell_1978a, Bell_1978b, Schlickeiser_2002, Schure_2012}.
If a charged particle gains sufficient energy it can then escape the supernovae shock upstream~\citep{Ohira_2019}. 
Lower energy charged particles remain trapped in the shock until the shock slows sufficiently for them to escape or for the shock to dissipate, at which point they are released into the surrounding medium. Except at the highest energies ($> 10^{7}$\,MeV) cosmic rays released as the shock dissipates dominate the time-integrated cosmic ray spectrum of a SNR~\citep{Ohira_2010, Caprioli_2010}. This cosmic ray spectrum is theoretically predicted~\citep{Bell_1978a, Longair_1994} to follow a power-law in particle kinetic energy $\ke{}$, $dN/d\ke{} \propto \ke^{-2}$, with an upper-cutoff of  $\sim 10^{9}$\,MeV~\citep[set by the SNR magnetic field strength and radius,][]{Lagage_1983, Schlickeiser_2002} and a lower-cutoff of $\sim 10^{-3}$\,MeV~\citep[set by the shock velocity, ][]{Sazonov_2015}.
Simulations find this spectrum to be insensitive to the environment of the supernovae~\citep{Sazonov_2015}. 
Since cosmic rays with energies above the predicted upper limit of $10^{9}$\,MeV are observed, there is strong evidence for a second source of cosmic rays, potentially active galactic nuclei (AGN) or gamma-ray bursts~\citep{Batista_2022}. However, in this study, we limit ourselves to considering only SNRs as sources of cosmic rays, as AGN are expected to be rare at the high redshifts of interest.

Simulations of DSA and Milky Way observations suggest cosmic rays carry away between 10 and 50 per cent of the initial kinetic energy of the shock~\citep{Drury_1989, Berezinskii_1990, Caprioli_2014}. The majority of this energy is in cosmic ray protons~\citep{Schlickeiser_2002, Leite_2017}, with per cent level portions of the energy in the form of alpha particles and electrons. Hence, we restrict our heating modelling and this review to considering cosmic ray protons only\footnote{Cosmic ray electrons are however important to consider when modelling SNR-produced radio backgrounds due to electrons producing a fraction  $(m_p/m_e)^2$ stronger synchrotron emission than protons~\citep{Jana_2019}.}.

\subsection{Escape of cosmic rays into the IGM} \label{ssec:cr_escape}

Based on the findings of numerical simulations~\citep{Kitayama_2005, Greif_2007, Whalen_2008}, \citet{Sazonov_2015} argued that Pop III SNRs forming in low-mass mini-halos ($\lesssim 10^7$\,\solm{}) would have escaped the virial radius due to their progenitor star having photoevaporated the gas out of the halo. The SNR then only forms a shock front and begins to suffer significant radiative energy losses once it caught up to the material previously ejected from the halo.
As a result, the SNR would not dissipate until it has propagated out to the distance of several virial radii of the host mini-halo which takes $1-10$~Myr. 
Since the recombination time estimated for the gas photoionized by the progenitor star is less than a few million years, \citet{Sazonov_2015} suggested that the cosmic rays produced by the SNR are directly released into the neutral IGM about the source mini-halo. For future ease of reference, we refer to this mechanism for cosmic rays reaching the IGM as {\it direct injection}.

In more massive halos, such as the present-day Milky Way, the stars are unable to photoevaporate the gas out to the virial radius. Hence, when supernovae occur in these more massive halos the SNR rapidly encounters the more dense gas causing the SNR to quickly lose its thermal energy radiatively and dissipate, releasing cosmic rays into the ISM. Observations in the Milky Way of the isotopic ratios of Beryllium~\citep{Schlickeiser_2002} suggest that cosmic rays diffuse out of the ISM into the surrounding IGM on a timescale of 10~Myr. \citet{Leite_2017} argued that since star-forming halos in the early universe are anticipated to be smaller and have weaker magnetic fields than the present-day Milky Way, a large proportion of cosmic rays would be able to diffuse out into the IGM In contrast to the previously discussed direct injection mechanism where cosmic rays are only able to escape from the lowest mass halos, this mechanism which we refer to as {\it diffusive escape} does not have an upper halo mass limit, and so cosmic rays can escape into the IGM from all star-forming halos.

The third mechanism by which cosmic rays can escape into the IGM is {\it galactic outflows}, studied by \citet{Samui_2018}. The authors demonstrated that 75 per cent of the IGM could be metal enriched and heated at $z = 8$ by galactic outflows driven by the pressure of confined cosmic rays if star formation occurs in molecular cooling halos. Due to the growth of these outflows, a significant portion of the cosmic ray energy is lost to adiabatic expansion. As in diffusive escape, there is no halo mass limit from which cosmic rays can be emitted by galactic outflows, though the spectrum of cosmic rays reaching the IGM will differ between the two mechanisms. Due to these similarities between the last two emission mechanisms,  we do not distinguish between them in this paper. All conclusions of our study that apply to cosmic rays diffusing out of halos should be also taken to apply to cosmic ray heating from galactic outflows.

\subsection{Energy-loss mechanisms of cosmic ray protons} \label{ssec:cr_energy_loss}

Once in the IGM, cosmic ray protons can lose their energy through numerous mechanisms including excitation, ionization, Coulomb energy exchange with charged particles, and particle production in collisions~\citep{Schlickeiser_2002}. However, when modelling cosmic ray heating, we only need to consider mechanisms contributing to efficient energy loss by cosmic ray protons in the energy range of interest. Hence, since in the early universe SNRs are believed to be the predominant cosmic ray sources, we limit our considerations to energies below $10^{9}$\,MeV, the highest energy SNRs can accelerate cosmic ray protons to. 

To determine whether an energy-loss mechanism for cosmic ray protons is efficient we compare it to the energy-loss rate due to the expansion of the universe.  As the universe expands the momentum of a cosmic ray proton decreases as the reciprocal of the scale factor. Consequently, the kinetic energy, $\ke$ (the energy potentially available for heating), of a proton decreases as
\begin{equation} \label{eqn:hubble_timescale_def}
    \left.\frac{d\ke{}}{dz}\right|_{\rm H} = \frac{\ke{}}{1+z} \left(2 - \frac{1}{1 +  (m_{\rm p} c^2/\ke{})}\right),
\end{equation}
where $m_{\rm p}$ is the mass of the proton. In this equation, the first term entails the non-relativistic limit, and the second is the relativistic correction. We can thus define the Hubble cooling timescale to be
\begin{equation}
    t_{\rm H} = \frac{- \ke{}}{\left.d\ke{}/dz\right|_{\rm H}  }\frac{dt}{dz},
\end{equation}
and for any other energy-loss mechanism (with placeholder label i) we can similarly define its energy-loss timescale as 
\begin{equation}
    t_{\rm i} = \frac{- \ke{}}{\left.d\ke{}/dz\right|_{\rm i}  }\frac{dt}{dz} = \frac{-\ke{}}{\left.d\ke{}/dt\right|_{\rm i}  }.
\end{equation}
The dominant energy-loss mechanism for a given value of $\ke{}$ at a given redshift $z$ is thus the mechanism with the lowest energy-loss time scale. In the case where Hubble cooling is dominant, the energy of the proton is principally lost to the expansion of the universe and will not contribute to the heating of the IGM.

Previous studies have shown point-like electromagnetic interactions (e.g. bremsstrahlung, inverse Compton scattering, and synchrotron) and collisions with CMB photons to be inefficient mechanisms for cosmic ray proton energy loss\footnote{These mechanisms have low cross-sections due to the small charge-to-mass ratio of the proton compared to the electron, and the low energy of CMB photons compared to the potential end product particle rest masses respectively.}~\citep{Schlickeiser_2002}. Furthermore, pion production by collisions between neutral hydrogen and cosmic ray protons is only efficient at energies above the range of interest to us, while resistive heating is only dominant in the highly ionized regions around galaxies invisible to the 21-cm signal~\citep{Yokoyama_2022, Yokoyama_2023}. Alongside Hubble cooling this leaves three energy-loss mechanisms to consider\footnote{A new mechanism by which cosmic rays can transfer heat to their surrounding medium has recently been proposed called \textit{self-discharge}~\citep{Ohira_2022}. Unfortunately, at this time estimates of the heating rate from this process are not available and so we do not consider it further in this work.}: excitation and ionization of neutral hydrogen atoms~\citep{Sazonov_2015, Jana_2018}, Coulomb interactions with free electrons~\citep{Leite_2017}, and Alfv\'en wave emission~\citep{Samui_2018, Bera_2022}.

If a cosmic ray proton collides with a neutral hydrogen atom it can directly ionize the atom. In each such interaction the cosmic ray proton losses $\approx 60$\,eV of energy to the liberated electron~\citep{Spitzer_1969, Sazonov_2015}, which can then ionize or excite surrounding hydrogen atoms in secondary collisions. The energy-loss rate of cosmic ray protons to this process is described by the Bethe–Bloch equation, which for protons with \mbox{$\ke{} < 8.4 \times 10^5$\,MeV} is well-approximated by~\citep{Schlickeiser_2002}
\begin{equation} \label{eqn:e_i_loss_rate}
\begin{split}
\left.\frac{d\ke{}}{dt}\right|_{\rm E\&I} \approx - &\left(1.82 \times 10^{-7} {\rm eV\,s}^{-1}{\rm\,cm}^{3}\right) x_{\rm HI} n_{\rm H} \\
& \times \left[1 + 0.0185 \ln (\beta) \Theta(\beta - \beta_0) \right] \frac{2 \beta^2}{\beta_0^3 + 2 \beta^3}.
\end{split}
\end{equation}
In the above equation, $\Theta$ is the Heaviside step function, $x_{\rm HI}$ the hydrogen neutral fraction, $n_{\rm H}$ the physical number density of hydrogen, $\beta = {\rm v}/c$ the normalized velocity of the proton, and $\beta_0 = 0.01$ the loss rate peak. Due to the small value of $\beta_0$, this process is anticipated to be the strongest for non-relativistic protons, and the strength of this mechanism will decrease as the IGM ionizes owing to its scaling with $x_{\rm HI}$.

Cosmic ray protons can also transfer energy directly to any free electrons in the IGM via Coulomb interactions, with this energy again being dissipated into heat in the IGM via the subsequent collisions of these energetic electrons. The equations to describe the rate at which cosmic ray protons lose energy to this mechanism were derived by \citet{Butler_1962}, \citet{Sivukhin_1965}, and \citet{Gould_1972} and are well approximated this time by~\citep{Schlickeiser_2002} 
\begin{equation}
\begin{split}
\left.\frac{d\ke{}}{dt}\right|_{\rm C} \approx &  \frac{-\left(3.1 \times 10^{-7} {\rm eV\,s}^{-1}{\rm\,cm}^{3}\right) x_{\rm e} n_{\rm b} \beta^2}{\left(T_{\rm e}/2.45 \times 10^9{\rm\,K}\right)^{3/2} + \beta^3},
\end{split}
\end{equation}
where $x_e$ is the electron fraction, $n_b$ the number density of baryons, and $T_{\rm e}$ is the temperature of the free electrons which we take to be the kinetic temperature of the IGM, $T_{\rm K}$. Hence, unlike excitation and ionization interactions, the efficiency of this mechanism increases as the IGM becomes more and more ionized.

Lastly, since cosmic rays are charged they gyrate along magnetic field lines $\mathbfit{B}$ generating Alfv\'en waves~\citep{Samui_2018} which carry away a portion of their energy. The dissipation of these Alfv\'en waves then transfers this energy to the IGM as heat at rate \mbox{$|\mathbfit{B} \cdot \nabla P_{\rm cr}|/ (4 \pi \rho)$} where $P_{\rm cr}$ is the pressure from cosmic rays and $\rho$ is the IGM physical density. Due to the dependence on $P_{\rm cr}$ this is a bulk effect, unlike the energy-loss mechanisms mentioned previously, which can be considered in terms of individual protons. Consequently, we cannot define an energy-loss timescale for this mechanism in the same way we have done above. However, \citet{Bera_2022} provides an estimate for when Alfv\'en wave dissipation is an efficient heating process by comparing the IGM heating timescale in this case ($\tau_{\rm A} = T_{\rm K} / \dot{T}_{\rm K}$) with the Hubble time ($\tau_{\rm H} = 1/H$), note the latter differs by a factor of order unity from $t_{\rm H}$ defined in equation~\eqref{eqn:hubble_timescale_def}. For the Planck 2018 best-fit $\Lambda$CDM cosmology~\citep{Planck_VI} this ratio becomes
\begin{equation}
\begin{split}
\frac{\tau_{\rm A}}{\tau_{\rm H}} \approx 0.14 & \left(\frac{1+z}{16}\right)^4 \left(\frac{T_{\rm K}}{\rm 10\,K} \right) \left(\frac{0.1{\rm\,nG}}{B_{\rm 0}}\right)\left(\frac{L}{0.01{\rm\,Mpc}} \right) \\
& \times \left(\frac{5 \times 10^{-5}{\rm\,eV\,cm}^{-3}}{U_{\rm cr}}\right).
\end{split}
\end{equation}
Here $B_0$ is the magnitude of the primordial IGM magnetic field seen today, $U_{\rm cr}$ the cosmic ray energy density, and $L$ the typical length-scale of cosmic ray pressure gradients taken by \citet{Bera_2022} to be the inter-halo distance. There is much uncertainty surrounding $B_{\rm 0}$ with upper bounds of $\sim 1$\,nG from CMB observations at 1\,cMpc scale~\citep{PLANCK_B} and lower limits from second-order perturbation theory of $\sim 10^{-11}$\,nG at 10\,ckpc ~\citep{Ichiki_2006}.
Recently there have also been attempts to measure the magnetic field in voids today with a disputed lower bound due to Blazar observations of $\sim 10^{-6}$\,nG~\citep{Tavecchio_2011}, and potential measurement of $\sim 10^{-7}$\,nG from gamma-ray bursts~\citep{Xia_2022}, though it is still unclear if the present-day void magnetic field is primordial in origin or has been enhanced by outflows from galaxies ~\citep{Samui_2018}. Thus, the reference value of $B_{\rm 0}$ adopted above is at the upper limits of the 11 orders of magnitude permitted range, with the majority of this range ($B_{\rm 0} \lesssim 0.01$\,nG) leading to inefficient Alfv\'en wave heating of the IGM for the redshift range of interest. Therefore, we elect to ignore Alfv\'en wave losses from cosmic rays in our modelling and instead revisit this mechanism briefly in section~\ref{sec:conclusions} to discuss how our results would be changed by its inclusion due to a high $B_{\rm 0}$. 
Note as the magnetic fields essential to DSA (see Section~\ref{ssec:cr_production}) are produced in the SNR shock itself by the Weibel instability, a weak background magnetic field would not prevent cosmic ray acceleration by the first supernovae~\citep{Ohira_2019}.

\begin{figure*}
	\includegraphics{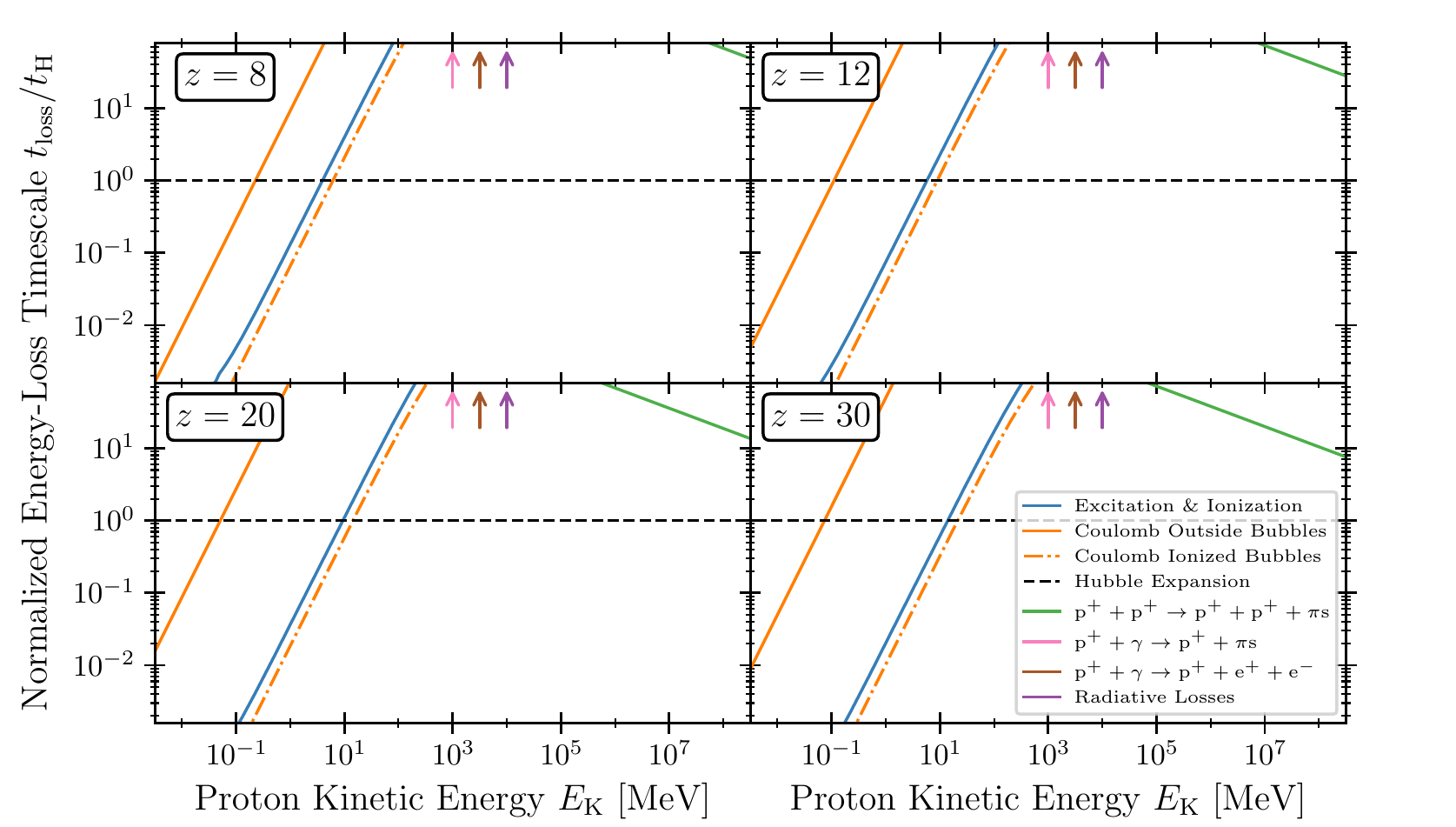}
    \caption{Comparison of energy-loss timescales  (normalized by the   Hubble cooling timescale to keep a consistent scale between redshifts) for cosmic ray protons at $z = 8$ (top left), $12$ (top right), $20$ (bottom left), and $30$ (bottom right). We consider excitation and ionization of neutral hydrogen (blue), Coulomb interactions with free electrons (orange), Hubble cooling (black), collisions with matter producing pions (green), collisions with photons producing pions (pink) or electron-positron pairs (brown), and radiative losses (purple). For Coulomb energy losses we distinguish between the IGM in an ionized bubble (dot-dashed) and outside of ionized bubbles (solid). Inside ionized bubbles we take $x_{\rm e} = 1$, and outside we use values of $0.0072$, $0.0014$, $0.0002$, and $0.0002$ in ascending redshift order from the output of the reference simulation (subsection~\ref{sssec:tomography}). $n_{\rm H}$ is taken as the cosmological average value. At high energies, $\ke{} \gtrsim 10$\,MeV, we find that Hubble cooling is the dominant mechanism (the lowest energy-loss timescale) both inside and outside ionized bubbles. For lower energies, Coulomb dominates in ionized bubbles, and excitation and ionization dominates outside of ionized bubbles. Pion-producing collisions with matter, photon collisions, and radiative losses remain subdominant for the energy range considered here, with the latter three being too inefficient to appear on the scale shown.     }
    \label{fig:cooling_timescales.pdf}
\end{figure*}

The relative importance of energy-loss mechanisms for cosmic ray protons can now be determined via a comparison of their energy-loss timescales $t_i$, with the lowest value corresponding to the dominant process. Such a comparison is shown in Fig.~\ref{fig:cooling_timescales.pdf} for $z = 8$, $12$, $20$, and $30$. In the figure, we normalize all energy-loss timescales via $t_{\rm H}$ to allow for easier comparison between redshifts. Since reionization is anticipated to be patchy~\citep{Choudhury_2022},  $x_{\rm e}$ (and with it efficiencies of some of the aforementioned mechanisms) is expected to vary greatly between different regions in our simulations. To demonstrate the importance of this effect for the cosmic ray heating, we show Coulomb heating timescales for both a fully ionized region (the case referred to as {\it Coulomb Ionized Bubbles} in the figure) and for the average value of  $x_{\rm e}$ outside of ionized bubbles taken from the reference simulation presented in  subsection~\ref{sssec:tomography} (the case referred to as {\it Coulomb Outside Bubbles}). Note that inside the ionized bubbles no excitation or ionization interactions occur, so the timescale depicted is that for cosmic rays outside of ionized bubbles. As we see from the figure, Hubble cooling dominates above $\ke{} \sim 30$\,MeV at all redshifts, suggesting higher energy relativistic cosmic rays do not contribute substantially to the IGM heating in agreement with the findings of \citet{Sazonov_2015}. At lower energies, outside ionized bubbles, excitation and ionization energy losses are dominant at all redshifts, while inside ionized bubbles Coulomb losses are dominant. Our findings at lower energies and low redshifts ($z\sim 8$) seemingly contradict the conclusions of \citet{Leite_2017}, who showed Coulomb energy-losses dominate excitation and ionization losses at lower energies towards the end of reionization. However, this is not a discrepancy as \citet{Leite_2017} used a globally averaged $x_{\rm e}$ while we have distinguished between fully-ionized and non-fully-ionized regions of the IGM. We reach the same conclusion as \citet{Leite_2017} when the globally averaged $x_{\rm e}$ is used. 

Since this paper is principally concerned with modelling the 21-cm signal of neutral hydrogen, we are primarily interested in heating outside of fully ionized bubbles. Therefore, the mechanisms of interest to us are excitation and ionization losses, and Hubble cooling. However, due to the similarity between the energy-loss timescales of Coulomb interactions inside ionized bubbles and excitation and ionization interactions outside of them, the resulting energy evolution equation for cosmic ray protons is anticipated to be a good model for cosmic rays propagating through ionized regions as well.

\subsection{Cosmic ray heating and ionization} \label{ssec:cr_heating}

Not all of the energy lost by cosmic rays, discussed above, is converted to the IGM heat. None of the energy lost to Hubble cooling is transferred to the IGM, and of the energy lost to excitation and ionization only a portion $f_{\rm heat}$ ultimately ends up as heat energy in the IGM with the rest lost to the ionization energy and the de-excitation spectral line emission of excited hydrogen atoms. \citet{Spitzer_1969} calculated the heat transfer in excitation and ionization interactions of cosmic ray protons for several different free electron fractions $x_{\rm e}$.

We note the value of $f_{\rm heat}$ is quite sensitive to small changes in $x_{\rm e}$, varying from 10\% to 53\% between $x_{\rm e}$ values of $10^{-4}$ and $0.5$. Therefore, while recent studies found cosmic rays do not contribute significantly to reionization~\citep{Sazonov_2015, Leite_2017}, ionization, including that from cosmic rays, may strongly impact the cosmic ray heating rate.  Thus, it should be considered and modelled self-consistently.

Per primary ionization of a neutral hydrogen atom, a cosmic ray proton losses $\approx 60$\,eV~\citep{Sazonov_2015}, of which the fraction $f_{\rm heat}$ contributes to heating the IGM. The primary ionization rate per baryon $\Lambda_{\rm cr}^{\rm primary} $ can then be found from the cosmic ray heating rate per baryon, $\epsilon_{\rm cr}$,
\begin{equation}
    \Lambda_{\rm cr}^{\rm primary} = \frac{1}{60{\rm\,eV} f_{\rm heat}} \epsilon_{\rm cr}.
\end{equation}
Primary electrons freed by cosmic ray interactions can then cause secondary ionizations in their subsequent collisions. The average number of secondary ionizations per primary ionization $\phi$ for various values of $x_{\rm e}$ were also calculated by \citet{Spitzer_1969}. Thus the full cosmic ray ionization rate per baryon is given by 
\begin{equation} \label{eqn:cr_ioniz_rate}
    \Lambda_{\rm cr} = \frac{1 + \phi}{60{\rm\,eV} f_{\rm heat}} \epsilon_{\rm cr}.
\end{equation}

\subsection{Propagation of cosmic rays through the IGM} \label{ssec:cr_propogation}

Since the early universe is not homogeneous alongside the rate of cosmic ray heating in the IGM, we must also consider its spatial distribution. Most previous studies of cosmic ray heating~\citep{Sazonov_2015, Leite_2017, Jana_2019, Bera_2022} have assumed the heating to be spatially uniform. This assumption is partially motivated by arguments that cosmic ray heating should reach inter-halo distances ($\gtrsim 0.1$cMpc at $z = 20$), based upon the Bohm diffusivity lower bound and Larmor radius of cosmic rays~\citep{Stacy_2007, Sazonov_2015}.

A more detailed study of the distribution of cosmic ray proton and the heating pattern around the source halos was undertaken by \citet{Jana_2018}, who modelled cosmic rays as diffusing out of their source halos with a constant diffusivity $D$. This is physically motivated by cosmic rays scattering from magnetic fields leading to them undergoing a random walk and is similar to models of cosmic ray propagation through the Milky Way ISM such as \textsc{GALPROP}~\citep{Strong_1998, Ptuskin_2012}. From their study~\citet{Jana_2018} conclude that cosmic ray heating around minihalos between $z = 10$ and $20$ was uniform, while it would be inhomogeneous around high-mass galaxies with low star formation rates. However, importantly, they conclude their results are not definitive due to the large uncertainties in the value of the cosmic ray diffusivity in the early universe. These uncertainties come about in part due to the large, up to 11 orders of magnitude, uncertainty in the IGM magnetic field discussed in the previous subsection, and due to the scaling of cosmic ray diffusivity with cosmic ray energy and magnetic field being unknown. Even if cosmic ray heating is uniform on the typical distance scale between minihalos as \citet{Jana_2018} tentatively concluded, this does not necessarily mean cosmic ray heating is uniform throughout the universe since at early times star-forming minihalos are highly clustered around rare overdensities~\citep{Barkana_2004, Barkana_2016}. Instead, it would imply cosmic ray heating is uniform on the scales on which minihalo number density is uniform.

The uncertainty surrounding the propagation of cosmic rays in the early universe IGM leads us to consider multiple different modes of cosmic ray propagation in our study, both to bracket the range of physically plausible scenarios and also to allow us to investigate the potential of the 21-cm signal to distinguish between these possibilities.

\section{Semi-Numerical 21-cm Signal Simulations}
\label{sec:21cm}

Having outlined the five necessary steps to modelling cosmic ray heating in the previous section, we introduce our 21-cm signal simulation code and detail the changes we make to include the new heating mechanism.

\subsection{Background theory and current simulation code}

21-cm cosmology aims to observe the absorption or emission at the 21-cm spectral line of atomic hydrogen~\citep{Furlanetto_2006}. The strength of the 21-cm signal is dependent on the number density of atoms in the hyperfine states $n_{\rm 1}$, where the magnetic moments of proton and electron are aligned, compared to the state $n_{\rm 0}$, where the magnetic moments are anti-aligned. Additionally, the signal depends on the background flux of photons at the 21-cm line, commonly quantified in terms of the radiation temperature $T_{\gamma}$. We can conveniently express the relative occupancy of the hyperfine states in terms of spin temperature,  $T_{\rm s}$,  via the use of the Boltzmann distribution~\citep{Scott_1990}
\begin{equation} \label{eqn:Ts_def}
    \frac{n_{\rm 1}}{n_{\rm 0}} = 3 \exp \left(- \frac{E_{\rm 21}}{k_{\rm B} T_{\rm s}} \right),
\end{equation}
where the factor of three comes from the three-fold degeneracy of the higher-energy states, and $E_{\rm 21} = 5.87$\,$\mu$eV (corresponding to $\nu_{\rm 21} = 1420$\,MHz) is the small energy difference between the aligned and anti-aligned states. The relative value of $T_{\rm s}$ to  $T_{\gamma}$ determines whether the signal is to be seen in emission \mbox{($T_{\rm s} > T_{\gamma}$)} or absorption \mbox{($T_{\rm s} < T_{\gamma}$)}. A more detailed treatment of the radiative transfer problem, accounting for the number density of hydrogen $n_{\textrm{H}}$, hydrogen neutral fraction $x_{\rm HI}$, and the gradual redshifting of photons into and out of the spectral line allows for the calculation of the differential 21-cm brightness temperature that would be seen today 
\begin{equation}
T_{\rm 21}  = \left(1 - e^{-\tau_{\rm 21}} \right)\frac{T_{\rm s} - T_\gamma}{1+ z},
\label{eqn:Tb_equation}
\end{equation}
formally defined as the difference in radiation temperature seen at $\nu = \nu_{\rm 21}/(1+z)$ due to the emission/absorption at the 21-cm line by the IGM at redshift $z$. In the above equation $\tau_{\rm 21}$ is the 21-cm optical depth given by
\begin{equation}
\tau_{\rm 21} = \frac{3}{32 \pi} \frac{h c^3 A_{\rm 10}}{k_{\rm B} \nu_{\rm 21}^2} \left[\frac{x_{\rm HI} n_{\rm{H}}}{(1+z)^2 (d{\rm v}_\parallel/dr_\parallel)} \right]\frac{1}{T_{\rm s}},
\label{eqn:21cm_optical_depth}
\end{equation}
where $A_\textrm{10} = 2.85 \times 10^{-15}$\,s$^{-1}$ is the spontaneous emission rate of the 21-cm transition, and $d{\rm v}_\parallel/dr_\parallel$ the proper velocity gradient along the line of sight including the Hubble flow. In summary, simulating the 21-cm signal requires determining $T_{\rm \gamma}$, $T_{\rm s}$, and $x_{\rm HI}$. In this study, we assume the background radiation temperature $T_{\gamma}$ to be just the CMB radiation, \mbox{$2.725 (1+z)$\,K}. We leave the simultaneous consideration of excess radio backgrounds~\citep{Feng_18, Ewall_18, Fialkov_2019,  Reis_2020} and cosmic ray heating to future investigations. 

The calculation of the spin temperature $T_{\rm s}$ of the IGM is complicated due to it being acted upon by three competing influences: collisional coupling to the kinetic temperature of the IGM ($T_{\rm K}$), radiative coupling to the background 21-cm radiation temperature ($T_{\gamma}$), and indirect coupling to $T_{\rm K}$ via the scattering of Lyman line photons, the Wouthuysen-Field effect~\citep[WF,][]{Wouthuysen_1952, Field_1958}. The efficiency of each of these processes at any given redshift is encapsulated by the corresponding coupling coefficient, $x_{\textrm{c}}$, $x_{\gamma}$, and $x_\alpha$, which are in turn derived from atomic physics and the intensity of the Lyman line radiation fields~\citep{Furlanetto_2006,Venumadhav_2018}. Once these coupling coefficients are found the spin temperature can be computed iteratively using 
\begin{equation} \label{eqn:Ts_couplings}
    \frac{1}{T_{\textrm{s}}} = \frac{x_\textrm{c} T_\textrm{K}^{-1} + x_{\gamma} T_\gamma^{-1} + x_\alpha T_\textrm{c}^{-1}}{x_\textrm{c} + x_{\gamma} + x_\alpha},
\end{equation}
where $T_\textrm{c}$ is the colour temperature of \lya{} photons~\citep{Barkana_2016}. 
At cosmic dawn, the first generation of stars is anticipated to have filled the early universe with a background of \lya{} photons causing the WF coupling to dominate the other processes (with $x_{\alpha} \gg x_{\gamma}, x_{\rm c}$) resulting in $T_{\rm s} \approx T_{\rm K}$. Hence before reionization when $x_{\rm HI} \approx 1$, but after cosmic dawn we anticipate (from equation~\ref{eqn:Tb_equation}) that the 21-cm signal should trace variation in the gas temperature of the IGM, making it particularly sensitive to cosmic ray heating along with other heating/cooling mechanisms. However, since our study spans a wider range of redshifts than those for which the aforementioned approximation is valid we always utilize equation~\ref{eqn:Ts_couplings} when calculating $T_{\rm s}$.

We thus also need to determine $T_{\rm K}$ and the Lyman line radiation fields to compute $T_{\rm s}$. $T_{\rm K}$ can be calculated by solving the ordinary differential equation describing the balance of heating and cooling influences on the IGM~\citep{Mesinger_2011}
\begin{equation}
\begin{split} \label{eqn:Tk_ode}
    \frac{dT_{\rm K}(z)}{dz} =& \frac{2}{3 k_{\rm B} (1 + x_{\rm e})} \frac{dt}{dz} \sum_{\rm p} \epsilon_{\rm p} \\ &+ \frac{2}{3} \frac{T_{\rm K}}{n_{\rm b}} \frac{dn_{\rm b}}{dz} - \frac{T_{\rm K}}{1 + x_{\rm e}}\frac{dx_{\rm e}}{dz} + \frac{2 T_{\rm K}}{1 + z}.
\end{split}
\end{equation}
The first term encompasses the heating/cooling rate per baryon, $\epsilon_{\rm p}$,  due to various astrophysical processes, in our simulations, we include X-ray heating~\citep{Fialkov_2014b}, Compton heating~\citep{Madau_1997}, \lya{} heating~\citep{Reis_2021}, and CMB heating~\citep{Venumadhav_2018, Fialkov_2019, Reis_2021}. Later we extend this list to include cosmic ray heating as well. Next, the second term represents heating due to structure formation, and the third term takes into account cooling due to the increasing number of particles caused by the reionization of the IGM. Lastly, the fourth term represents the adiabatic cooling of the gas due to the expansion of the universe.

To compute the Lyman line radiation fields, and thus $x_{\alpha}$, we follow the methodology of \citet{Reis_2021}, an extension of \citet{Barkana_2005b} and \citet{Fialkov_2014}.
Hence, we also assume the Wouthuysen-Field effect is dominated by \lya{} photons due to such photons scattering orders of magnitude more times than higher Lyman line photons~\citep{Furlanetto_2006}.
As a result, in this model, the WF coupling is taken to simply be directly proportional to the \lya{} line radiation field $J_\alpha$~\citep{Madau_1997}
\begin{equation}~\label{eqn:xalpha}
    x_{\alpha} = \frac{1}{A_{\rm 10} T_\gamma} \frac{16 \pi^2 E_{\rm 21} e^2 f_{\alpha}}{27 m_{\rm e} c k_{\rm B}} J_\alpha,
\end{equation}
where $f_{\alpha} = 0.4162$ is the oscillator strength of the \lya{} transition.
Two separate contributions to $J_\alpha$ are modelled, photons that directly redshift into \lya{} and photons produced in the cascading decays of neutral hydrogen excited by higher Lyman line photons. 
For the latter, it is assumed that photons emitted between Ly\,$\beta$ and the Lyman limit travel on radial geodesics away from sources until they redshift into the Lyman line below their emission frequency and cascade into a \lya{} photon with a line dependant recycling fraction. 
Whereas the paths of photons emitted between \lya{} and Ly\,$\beta$ that directly redshift into \lya{}, are treated as random walks due to their many scatterings in the tails of the \lya{} line. 
In the simulation, both radiative transfer processes are implemented using isotropic window functions.
These window functions are analytic spherical shells for the cascading photons and fits to Monto Carlo simulations for directly redshifting photons. 
By convolving the window functions with emissivity fields, which are in turn derived from the past star formation rate and stellar population, $J_\alpha$ and $x_{\alpha}$ can be calculated efficiently throughout the simulation box.

Finally, we need to compute the neutral fraction $x_{\rm HI}$ of the IGM, for which we adopt an approach similar to~\citet{Mesinger_2011}.
We use an excursion set-based formalism~\citep{Furlanetto_2004} to identify regions that have been fully ionized ($x_{\rm HI} = 0$) by galactic UV emission, which is anticipated to be the dominant mechanism driving reionization~\citep{Yajima_2011, Yajima_2014, Wise_2014, Ma_2020}.
A region is considered fully ionized if there exists a spherical volume of radius $R$ centred on that region in which the time-integrated ionizing UV emission exceeds the effective number of neutral atoms within said volume. 
Hence, for an effective galactic ionization efficiency per baryon $\zeta$, and collapse fraction of baryons into galaxies $f_{\rm coll}(\mathbf{x}, R)$ averaged over a volume of radius $R$ centered at $\mathbf{x}$, the point $\mathbf{x}$ is considered ionized if
\begin{equation}~\label{eqn:crude_reio}
    \exists R < R_{\rm mfp},\quad \textrm{s.t.\ } \quad \zeta f_{\rm coll}(\mathbf{x}, R) \geq 1.
\end{equation}
Where $R_{\rm mfp}$ is some maximum radius UV photons can travel to, normally set to the mean free path of ionizing photons in the ionized IGM at the end of reionization~\citep{Furlanetto_2005}, hence the choice of notation.
However, such an approach on its own would ignore the residual ionization leftover from recombination and any ionization from other sources such as X-rays~\citep{Pacucci_2014} or the cosmic rays we describe in this paper. 
Hence, we use a modified version of equation~\eqref{eqn:crude_reio} introduced by \citet{Mesinger_2013} to take into account the region already being partially ionized by these other mechanisms. With $\mathbf{x}$ considered ionized if
\begin{equation}~\label{eqn:better_reio}
    \exists R < R_{\rm mfp},\quad \textrm{s.t.\ } \quad \zeta f_{\rm coll}(\mathbf{x}, R) \geq 1 - x_{\rm e, oth}(\mathbf{x}, R) ,
\end{equation}
where $x_{\rm e, oth}(\mathbf{x}, R)$ is the ionization fraction of the IGM when not including the UV contribution, averaged over the same sphere as $f_{\rm coll}(\mathbf{x}, R)$. 
$x_{\rm e, oth}(\mathbf{x})$ in turn is determined by solving the ionization differential equation
\begin{equation} \label{eqn:xe_ode}
    \frac{dx_{\rm e, oth}(\mathbfit{x}, z)}{dz} = \frac{dt}{dz} \left(\Lambda_{\rm ion} - n_{\rm H} \alpha_{\rm B} x_{\rm e, oth}^2\right),
\end{equation}
with the first term encapsulating ionization at a rate of $\Lambda_{\rm ion}$ per baryon from non-UV sources (e.g.\ X-rays and cosmic rays), and the second term modelling type-B recombination~\citep{Furlanetto_2006} with recombination coefficient $\alpha_{\rm B} = 2.6 \times 10^{-13} (T_{\rm K}/10^4{\rm\,K})^{-0.7}$\,cm$^3$\,s$^{-1}$. 
It is assumed in our simulations that the ionization fraction of hydrogen and helium is the same, as a result helium double ionization is not modelled. 
This reionization model is hence built on the assumption that ionization from UV is isolated to fully ionized bubbles, whereas other sources of ionization can travel into the neutral IGM and give rise to partial ionization. 
In practice, such bubbles may be smaller than the resolution of a simulation, which if not accounted for could lead to erroneous 21-cm signal predictions.
To mitigate this issue regions that are not fully ionized are modelled as a two-phased medium, one fully ionized and one ionized to $x_{\rm e, oth}(\mathbf{x})$, an approximation that was validated by \citet{Zahn_2011} against radiative transfer simulations. 
The relative proportion of the region taken up by each phase being given by the region's overall neutral fraction
\begin{equation} \label{eqn:partial_ion}
    x_{\rm HI}(\mathbf{x}) = 1 - \zeta f_{\rm coll}(\mathbf{x}) - x_{\rm e, oth}(\mathbf{x}).
\end{equation}
Combined this multi-stage approach to ionization allows for the modelling of the propagation of fully ionized bubbles on both sub-resolution and resolved scales, while also accounting for the evolution of the ionization fraction of the mostly-neutral IGM outside these bubbles.

The above equations combined with a parameterized prescription of star formation rate and stellar/galactic emission form the foundations of our semi-numerical 21-cm signal simulation code~\citep[e.g.,][]{Visbal_2012, Fialkov_2014, Cohen_2016, Reis_2021}. The simulation starts from cosmological initial conditions for overdensity and baryon-dark matter relative velocity computed using \textsc{CAMB}~\citep{Lewis_1999, Lewis_2002, Lewis_2011}, and initial conditions for the gas temperature and residual ionization fractions computed using \textsc{RECFAST} \citep{RECFAST}. For the purposes of this study, all these fields are created on a 128$^3$ grid of cells, each cell being a 3\,cMpc sided cube. These initial conditions are then evolved forward in time, at each step an analytic prescription being used to compute the expected halo mass distribution within each simulation cell~\citep{Barkana_2004, Fialkov_2012}. The code then employs the star formation prescription of \citet{Magg_2022} to translate these halo mass distributions into Pop III and Pop II star formation rates for each cell and in turn into an emission rate for various types of electromagnetic radiation. Our radiative transfer model then propagates the radio, Lyman band, and X-ray radiation fields via pre-computed window functions\footnote{The accuracy of using the window functions of \citet{Reis_2021} for \lya{} radiative transfer has been recently explored in \citet{Semelin_2023} through a comparison to full radiative transfer models. The window functions approach is found to perform well when not including peculiar velocities, but errors in the 21-cm power spectrum are seen when modelling of peculiar velocities are included. We are not aware of any similar comparative studies assessing the accuracy of the window functions approach for radio or X-ray radiative transfer.}. With the radiation fields in each cell, equations~\eqref{eqn:Tb_equation}, \eqref{eqn:Ts_couplings}, \eqref{eqn:Tk_ode}, \eqref{eqn:xalpha}, \eqref{eqn:better_reio}, \eqref{eqn:xe_ode} and \eqref{eqn:partial_ion} are then solved simultaneously to determine the expected 21-cm signal. Ultimately the simulations provide full tomographic maps of $T_{\rm 21}$ at each redshift step from which other 21-cm observables can be computed.

Due to the uncertainty surrounding the universe between recombination and reionization, many aspects of our semi-numerical simulations are parameterized as exact values are not known. Except when stated otherwise in this paper we use the parameters listed in Table~\ref{tab:sim_default_parameters}. 

\begin{table}
 \caption{Default values of 21-cm signal simulation parameters used throughout this work. If a simulation uses parameter values other than those listed here those that differ and their values will be explicitly stated.}
 \label{tab:sim_default_parameters}
 \begin{tabular*}{\columnwidth}{lll}
  \hline
  Parameter & Value & Description\\
  \hline
  $f_{*,\rm II}$ & 0.05 & Pop II star formation efficiency \\
  $f_{*,\rm III}$ & 0.02 & Pop III star formation efficiency \\
  $t_{\rm recov}$ & 30\,Myr & Recovery time of star-forming halos \\
  $\zeta$ & 15 & Effective galactic ionization efficiency\\
  $R_{\rm mfp}$ & 50\,cMpc & Mean free path of ionizing photons \\
  $f_{\rm X}$ & 0 & Galactic X-ray emission efficiency \\
  $f_{\rm rad}$ & 0 & Galactic radio emission efficiency \\
  $p_{\rm LW}$ & 0.75 & LW feedback delay~\citep{Fialkov_2013}\\
  Pop III $\alpha$  & Log-flat & Exponent of Pop~III IMF\\
  Pop III $M_{\rm min}$& 2\,M$_{\odot}$ & Minimum Pop~III mass\\
  Pop III $M_{\rm max}$ & 180\,M$_{\odot}$ & Maximum Pop~III mass\\
  \hline
 \end{tabular*}
\end{table}

Our ionization model is described by two of these parameters $\zeta$ the effective galactic ionization efficiency per baryon, and $R_{\rm mfp}$ the mean free path of ionizing photons in the ionized IGM at the end of reionization. While $R_{\rm mfp}$ has a direct physical interpretation $\zeta$ is phenomenological, encapsulating the star formation efficiency of ionizing sources, the ionizing UV escape fraction, the stellar UV emissivity, and the reduction in effective ionization efficiency due to recombinations and clumping of the IGM~\citep{Furlanetto_2004}. In an upcoming paper, we refine this reionization model by modelling the individual physically-interpretable components of $\zeta$. Since we observe the impacts of cosmic ray heating on the 21-cm signal to be strongest prior to $z = 10$ we find varying these reionization parameters has no impact on our qualitative conclusions. Hence throughout this paper, we fix the values of these two parameters. With $\zeta$ set to 15, to recover optical depths to the CMB of $\tau \approx 0.06$ consistent with the Planck~2018 measurements~\citep{Planck_VI}, and $R_{\rm mfp}$ fixed to 50\,cMpc, motivated by the theoretical expectation of a mean free path of ionizing photons in the ionized IGM of $70$\,cMpc~\citep{Wyithe_2004} at $z = 6$. 

For all of our simulations, we enable baryon-dark matter relative velocities~\citep{Fialkov_2012}, \lya{} heating~\citep{Reis_2021}, CMB heating~\citep{Venumadhav_2018}, Pop II star formation rate suppression in low-mass halos~\citep{Fialkov_2013}, Ly\,$\alpha$ multiple scattering~\citep{Reis_2021}, photoheating feedback~\citep{Cohen_2016}, and model the non-instantaneous emission of Pop~III stars~\citep{Gessey-Jones_2022}. In this work, we do not enable Poisson fluctuations of star-forming halos~\citep{Reis_2022}. We derive the Pop~III star Lyman band and Lyman-Werner band emissivities self consistently from the power-law IMF specified in Table~\ref{tab:sim_default_parameters}. The ionizing UV emissivity and X-ray emissivity of Pop III stars are not currently derived from the IMF but will be in future studies~(Gessey-Jones et al.\ in preparation; Liu et al.\ in preparation).

Unlike future experiments such as SKA2-LOW~\citep{Koopmans_2015}, which are expected to make full tomographic maps of $T_{\rm 21}$, experimental efforts so far have concentrated on measuring either the sky-average of the signal $\langle T_{\rm 21} (z) \rangle$~\citep{SCI-HI, BIGHORNS, EDGES, LEDA, PRIZM,  DAPPER, SARAS3, REACH} or the power spectrum of its spatial variations~\citep{LWA,  PAPER, MWA,  LEDA, LOFAR, NenuFAR, HERA}. Throughout this paper, we use the $\Delta^2$ power spectrum convention, 
\begin{equation}
    \left\langle \tilde{T}_{\rm 21} \left(\mathbfit{k},z\right) \tilde{T}^*_{\rm 21} \left(\mathbfit{k}',z\right) \right\rangle = (2 \pi)^3 \delta^D\left(\mathbfit{k} - \mathbfit{k}'\right)  \frac{2 \pi^2 }{k^3}\Delta ^2(k,z),
\end{equation}
where $\mathbfit{k}$ is the comoving wavevector and $\delta^D$ the Dirac delta-function.  Given the experimental interest, we concentrate our study on cosmic ray heating signatures in the global 21-cm signal, power spectrum and tomographic maps. For our analysis, we include redshift space distortions when computing the 21-cm global signal and power spectrum but not when depicting 21-cm tomographic maps.

\subsection{Updating Lyman-Werner feedback prescription}
\label{subsec:LW}

In addition to introducing cosmic ray heating as described in the subsequent subsection, we make a further modification to our semi-numerical simulation code. Previous versions of our simulation code utilized the fitting formula from \citet{Fialkov_2012} to model the minimum halo mass required for star formation via molecular cooling, this formula included modelling of the increase of this mass threshold due to the Lyman-Werner feedback and baryon-dark matter relative velocities. \citet{Munoz_2022}, \citet{Nebrin_2023}, and \citet{Hegde_2023} recently highlighted the \citet{Fialkov_2012} formula does not account for the mitigation of the LW feedback by ${\rm H}_{\rm 2}$ self-shielding, and consequently over-predicts the strength of the LW feedback compared to what is seen in the numerical simulations of \citet{Schauer_2021} and \citet{Kulkarni_2021}. Hence, we adopt the fitting formula of \citet{Munoz_2022} which has been calibrated to the aforementioned recent numerical simulations. 

\citet{Munoz_2022}, motivated by \citet{Kulkarni_2021}, assume that the increase in the mass threshold due to the LW feedback and baryon-dark matter relative velocities can be each accounted for by factors $f_{\rm LW}$ and $f_{{\rm v}_{\rm cb}}$ which combine multiplicatively. By  comparison to numerical simulations \citet{Munoz_2022} find a best-fit model of 
\begin{equation}
    f_{\rm LW} = 1 + 2.0 (J_{\rm 21})^{0.6},
\end{equation}
where $J_{\rm 21}$ is the LW band intensity normalized to $10^{-21}$\,erg\,s$^{-1}$\,cm$^{-2}$\,sr$^{1}$\,Hz$^{-1}$, and
\begin{equation}
    f_{v_{\rm cb}} = \left(1 + \frac{{\rm v}_{\rm cb}}{{\rm v}_{\rm rms}} \right)^{1.8},
\end{equation}
where ${\rm v}_{\rm rms}$, is the root mean square velocity of the baryon-dark matter relative velocities at the relevant redshift. 
The critical halo mass for molecular cooling star formation is then given by
\begin{equation}
    M_{\rm mol, crit} = M_{z = 20} f_{\rm LW} f_{{\rm v}_{\rm cb}}  \left(\frac{1+z}{21}  \right)^{-3/2} 
\end{equation}
where the redshift dependant term is motivated by the simulation findings of \citet{Kulkarni_2021} and theoretical predictions of \citet{Tegmark_1997}. In this study we take \mbox{$M_{z = 20} = 5.8 \times 10^5$\,\solm}. This value was chosen due to it being $1\sigma$ below the mean mass at which molecular cooling halos start forming stars in the $J_{\rm 21} = 0$, ${\rm v}_{\rm bc} = 0$ simulation of \citet{Schauer_2021}. We use $1\sigma$ below the mean (instead of using the minimum or average mass) to account for the fact that we have a sharp cutoff with mass in halos that are Pop~III star-forming (as opposed to the gradual transition found in the simulation). Therefore, using the minimum or average would lead us to over-predict or under-predict the star formation rate respectively. 
Additionally, in our simulations, we correct for the delayed response between molecular cooling star formation and the exponentially growing $J_{\rm 21}$ by employing the methodology of \citet{Fialkov_2013}. 
Instead of using the $J_{\rm 21}$ field of the current simulation time step in the feedback calculation, we use the field from when the universe was a fraction $p_{\rm LW} = 0.75$ of its age at that timestep.
$p_{\rm LW}$ can be varied from this fiducial value to simulate stronger and weaker than expected LW feedback.

The above feedback prescription allows us to model the minimum halo mass required for star formation due to molecular cooling. However, star formation can also occur via atomic cooling or, if a previous generation of stars has metal-enriched a halo, metal cooling. To account for atomic cooling we take the minimum halo mass for Pop~III stars to form as the lower of the molecular cooling threshold $M_{\rm mol, crit}$ and the atomic cooling threshold $M_{\rm atm, crit}$~\citep{Fialkov_2013}, which is not impacted by the LW feedback. Since we follow the star formation prescription of \citet{Magg_2022} we model metal-cooling Pop~II star formation separately to Pop~III stars. This prescription is based upon fits to \textsc{A-SLOTH}~\citep{ASLOTH} merger tree models, and assumes Pop~II star formation occurs rapidly due to metal-cooling once a halo has had time to recover from the ejection of material by the supernova of the first generation of stars. External metal enrichment of halos is not included in the model as the authors of that study found including such a mechanism had little impact on the Pop~II and Pop~III star formation rates they derived. Since Pop~III and Pop~II star formation are triggered by distinct cooling mechanisms our simulations take separate star formation efficiencies $f_{*,\rm III}$ and $f_{*,\rm II}$ for these processes.

\subsection{Our cosmic ray heating model} \label{ssec:cr_heating_summary}

We now describe the model for cosmic ray heating we have integrated into our semi-numerical simulations. The model was intentionally developed to be versatile and include free parameters in order  to encapsulate the range of different approaches of previous studies and allow for the uncertainties in the underlying physics discussed in section~\ref{sec:cr_heating}.

\subsubsection{Cosmic ray sources}~\label{sssec:cr_sources}

In previous studies, \citet{Sazonov_2015} consider only Pop III stars in low-mass halos as cosmic ray sources, whereas \citet{Leite_2017} consider all Pop II star-forming halos as sources, and \citet{Bera_2022} consider both. We encompass all these possibilities in our model. Furthermore, to account for the fact that in some models cosmic rays can only efficiently escape from low-mass halos ($\lesssim 10^7$\,\solm{}), we introduce a parameter $M_{\rm cr}^{\rm max}$, and restrict the source of cosmic rays in our simulation to a user-specified stellar population(s) in halos of mass less than  $M_{\rm cr}^{\rm max}$. 
In our simulations, we thus define the cosmic ray emitting star formation rate density field, SFRD$_{\rm cr}(M_{\rm cr}^{\rm max})$, as the star formation rate density (in each simulation cell) of objects that emit cosmic rays and are hosted in halos of mass less than  $M_{\rm cr}^{\rm max}$.

Due to the uncertainties in the fraction of stars that underwent supernovae, supernova kinetic energy yields, the conversion rate of this energy into cosmic rays, and the escape fraction into the IGM,  we convert SFRD$_{\rm cr}(M_{\rm cr}^{\rm max})$ into a cosmic ray energy injection rate via a combined efficiency factor $\eta_{\rm cr}$.  In other words,  $\eta_{\rm cr}$ is formally defined to be the proportionality constant between star formation rate and the rate of cosmic ray energy injected into the IGM. Since massive stars that undergo supernovae have short lives on cosmological timescales, we ignore the time delay between star formation and supernovae in our modelling. Motivated by theoretical predictions, we take the spectrum of cosmic rays injected into the IGM as a power-law
\begin{equation}
    \frac{dN}{d\ke{}} \propto \ke^{\alpha_{\rm cr}} \qquad [\kemin{} \leq \ke{} \leq \kemax{}],
\end{equation}
with exponent $\alpha_{\rm cr}$, lower kinetic energy cutoff $\kemin{}$ and upper kinetic energy  cutoff $\kemax{}$. We keep $\alpha_{\rm cr}$, $\kemin$ and $\kemax$ as free parameters to account for the potential difference between the spectrum of cosmic rays produced at the shock front and that of the rays injected into the IGM. The parameterization also allows us to explore the sensitivity of cosmic ray heating to the cosmic ray spectrum, which was previously found to be significant for variations in $\kemin$ \citep{Sazonov_2015} and $\alpha_{\rm cr}$~\citep{Leite_2017}\footnote{Note a momentum power-law spectrum was used in this study rather than the kinetic energy power-law employed in this paper.}. Thus we compute the injection rate of cosmic rays into the IGM per time interval $dt$ and volume $dV$ at cell $\textbfit{x}$ to be
\begin{equation} \label{eq:injected_spectrum_model}
    \frac{d^3N(\mathbfit{x}, \ke{}, z)}{d\ke{} dt dV} = \frac{\ke^{\alpha_{\rm cr}}}{\mathcal{N}(\alpha_{\rm cr})}\eta_{\rm cr} {\rm SFRD}_{\rm cr}(\mathbfit{x}; M_{\rm cr}^{\rm max}),
\end{equation}
for $\kemin{} \leq \ke{} \leq \kemax{}$, where the normalization factor is defined to be
\begin{equation}
    \mathcal{N}(\alpha_{\rm cr}) = \begin{cases}
    \ln(\kemax{}) - \ln(\kemin{}), & \alpha = -2,\\
    \left(\kemax^{2+\alpha_{\rm cr}} - \kemin^{2+\alpha_{\rm cr}}\right)/\left(2+\alpha_{\rm cr}\right), & \text{otherwise}.
  \end{cases}
\end{equation}

\subsubsection{Energy-loss and propagation}~\label{sssec:propogation}

With our source term specified, we now turn our attention to how cosmic ray protons behave inside the IGM.

By combining equations~\eqref{eqn:hubble_timescale_def} and~\eqref{eqn:e_i_loss_rate}, accounting for the energy losses to Hubble cooling and excitation and ionization respectively, we can construct a differential equation that describes the evolution of a cosmic ray proton energy in the neutral IGM 
\begin{equation} \label{eqn:cr_energy_ode}
\begin{split}
\frac{d\ke{}}{dz} =&  \left.\frac{d\ke{}}{dz}\right|_{\rm H} + \left.\frac{d\ke{}}{dt}\right|_{\rm E\&I} \frac{dt}{dz} \\
\approx& \frac{1}{1+z} \left(\ke{} + \frac{\ke{}}{1 +  \ke{} /(m_{\rm p} c^2)}\right) - \left(1.82 \times 10^{-7} {\rm eV\,s}^{-1}{\rm\,cm}^{3}\right)  \\
& \times x_{\rm HI} n_{\rm H}  \left[1 + 0.0185 \ln (\beta) \Theta(\beta - \beta_0) \right] \frac{2 \beta^2}{\beta_0^3 + 2 \beta^3} \frac{dt}{dz},
\end{split}
\end{equation}
which we can solve numerically to find the kinetic energy of a cosmic ray emitted at redshift $z_{\rm 0}$ with initial kinetic energy $\kezero{}$.

 The above equation describes the temporal evolution of our cosmic ray distribution but we also need the spatial distribution. Given the large uncertainty in how cosmic rays move through the primordial IGM, we consider several alternative models for the spatial distribution of cosmic rays  about their source, $P(\mathbfit{x}, z; \kezero{}, z_{\rm 0})$.

\begin{figure*}
    \centering
    \begin{subfigure}[t]{0.49\textwidth}
        \centering
	\includegraphics{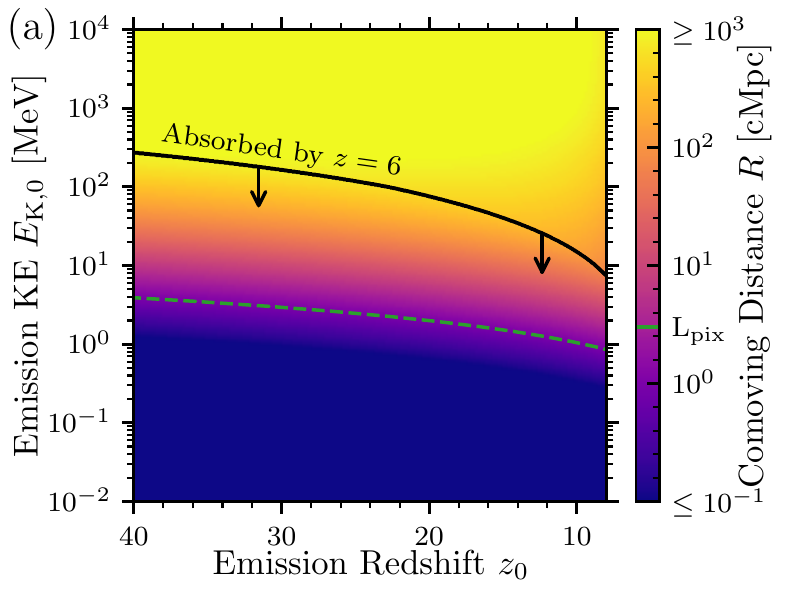}
        \phantomsubcaption
        \label{fig:comoving_path_lengths}
    \end{subfigure}
    \hfill
    \begin{subfigure}[t]{0.49\textwidth}
        \centering
	\includegraphics{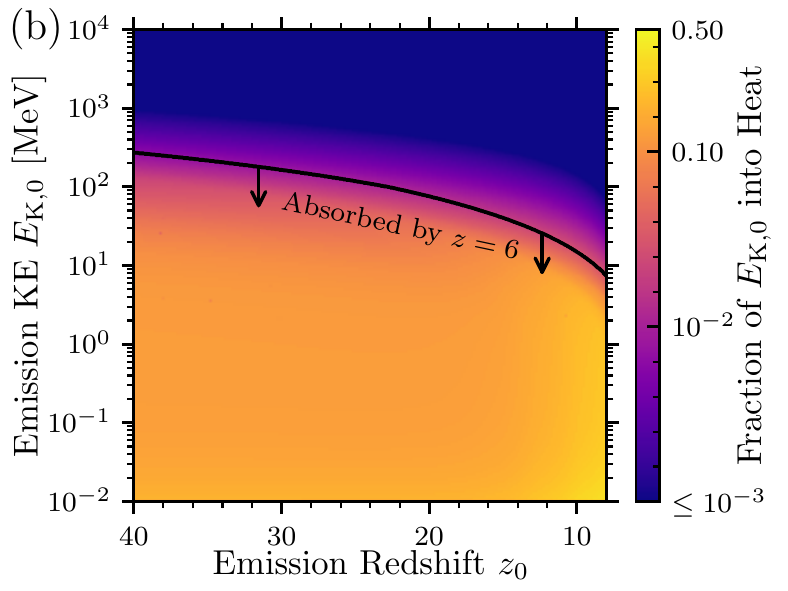}
        \phantomsubcaption
        \label{fig:fraction_heating}
    \end{subfigure}
    \caption{Comoving path length (panel a) of a cosmic ray proton at $z = 6$ emitted at $z_{\rm 0}$ with kinetic energy $\kezero{}$, and the fraction of its initial kinetic energy that is converted into the IGM heat (panel b). Here we have assumed the cosmic rays are moving in partially ionized IGM, with cosmic mean density and $x_{\rm e}$ taken from the reference simulation (subsection~\ref{sssec:tomography}). The black line indicates the $\kezero{}$ below which a cosmic ray is fully absorbed before $z = 6$. In panel a, the side length of one of our simulation cells is indicated as a green dashed contour for a reference, showing that cosmic ray protons emitted with energies below $\sim 2$\,MeV should not be able to escape their simulation cell of origin. Free-streaming cosmic rays emitted above $\sim 2$\, MeV but below $\sim 200$\,MeV can travel between cells but are still absorbed before $z = 6$, while protons with  $\kezero{} > 200$\,MeV are never absorbed. As shown in panel b, low energy cosmic rays ($\kezero{} < 10$\,MeV) deposit a significant portion ($> 10$ per cent) of their initial kinetic energy as heat, with this fraction increasing with redshift due to the increase in $f_{\rm heat}$ with $x_{\rm e}$. Conversely, at  $\kezero{} > 1000$\,MeV less than $0.1$ per cent of a cosmic ray kinetic energy becomes heat in the IGM, at least by $z = 6$.  } 
\end{figure*}

Physically, the furthest a cosmic ray could have reached from its source is set by its comoving path length $R$, with the cosmic ray free-streaming on a straight line path from its source. This can be calculated by integrating
\begin{equation} \label{eqn:cosmic_ray_eom}
    \frac{dR}{dt} = c(1+z)\sqrt{1 - \left(1 + \frac{\ke{}}{m_{\rm p} c^2} \right)^{-2} },
\end{equation}
where $\ke{}$ is in turn found by solving equation~\eqref{eqn:cr_energy_ode}. Such a {\it free-streaming} scenario could occur if the IGM magnetic field is very weak and so magnetic scattering of cosmic rays is rare. In this model $P(\mathbfit{x}, z; \kezero{}, z_{\rm 0})$ would become a shell window function at a comoving distance $R$ from the origin. To gauge the length-scale of this free-streaming propagation, which gives us the maximum possible range of cosmic ray heating, we calculate the comoving path length of protons when they are absorbed $R_{\rm abs}$. If the protons are not absorbed by $z= 6$ we calculate their distance (at $z=6$) to the emission source. The resulting values are depicted in Fig.~\ref{fig:comoving_path_lengths}. We find lower energy cosmic rays with initial kinetic energy $\lesssim 2$\,MeV travel less than the length of one of our simulation cells, $L_{\rm pix} = 3$\,cMpc. Above this energy threshold but below $\sim 200$\,MeV cosmic rays travel at least a simulation cell length and up to 1000\,cMpc while still being absorbed before $z = 6$. Therefore, we find that in principle cosmic rays can travel far from the source distributing the energy as heat over large cosmological scales (of the order of a few hundred cMpc). Cosmic rays with higher initial energies ($\gtrsim 200$\,MeV) are not absorbed by $z = 6$. From our comparison of energy-loss timescales in subsection~\ref{ssec:cr_energy_loss}, we know for cosmic rays with $\ke{} > 200$\,MeV their energy losses are dominated by Hubble cooling. Hence, while these highest-energy cosmic rays travel large ($> 1$\,cGpc) cosmological distances they are not anticipated to contribute to cosmic ray heating due to their energy being lost to the expansion of the universe.

For cosmic ray spectrum with the theoretically predicted exponent of $\alpha_{\rm cr} = 2$ the injected cosmic ray energy is split evenly over log initial kinetic energy. Hence, the low energy cosmic rays with \mbox{$\ke{} \lesssim 2$\,MeV} are anticipated to contain a significant fraction~\citep[$30$\,per cent, using the $\kemin{}$ and $\kemax{}$ of][]{Sazonov_2015} of the total injected cosmic ray energy. Since these cosmic rays can only travel short distances on cosmological scales we would thus anticipate that even in this limiting case of free-streaming a substantial portion of cosmic ray heating occurs within a few cMpc of the source halo, and thus cosmic ray heating is likely strongly clustered around star-forming halos. This localized nature of cosmic ray heating would in turn suggest that the globally averaged models used in previous works are insufficient to fully model the impact of cosmic ray heating on the 21-cm signal. For comparison with other studies, we will also consider a spatially uniform case of cosmic ray heating with $P(\mathbfit{x}, z; \kezero{}, z_{\rm 0})$ being a constant. However, we note that this is not a physically plausible model of cosmic ray propagation due to its acausal nature. 

The opposite logical extreme to free-streaming would be for cosmic rays to not escape their parent halos.
This would correspond to no cosmic ray heating of the neutral IGM, and hence no impact on the 21-cm signal (easily modelled by setting $\eta_{\rm cr} = 0$). 
However, cosmic rays can escape the Milky Way~\citep{Schlickeiser_2002}, a much larger and thus harder to escape galaxy than those expected at high redshifts. 
In addition, \citet{Sazonov_2015} argued that the Bohm lower diffusivity bound showed cosmic ray heating should extend to distances greater than the average separation between minihalos at $z \leq 20$. 
Similarly, detailed simulations by \citet{Jana_2018} of the diffusion of cosmic rays out of individual halos led them to conclude that cosmic ray heating from minihalos at $z = 10$ and $z = 20$ is homogeneous on inter-halo scales, even when the diffusivity of cosmic rays is assumed to be two orders of magnitude less than is seen in the Milky Way today. 
The physical expectation is thus that even in the most confined scenarios some heating cosmic rays will escape into the IGM.

Previous studies would thus indicate the minimum physically plausible range of cosmic ray heating is comparable to the inter-halo distance at $z < 20$.
For our model of star formation, we find each cell of our simulation should contain several star-forming halos at $z \lesssim 20$~\citep{Reis_2022}.
Hence for the epochs during which the 21-cm signal is driven by heating, the cosmic ray minimum range is below the resolution of our simulations.
As a result, in our shortest plausible range model, hereby called the {\it locally-confined} heating model, cosmic ray heating should not travel between neighbouring simulation cells, e.g.\ $P(\mathbfit{x}, z; \kezero{}, z_{\rm 0})$ is 1 in the origin cell and 0 elsewhere.

In the locally-confined heating model the distribution of cosmic ray heating thus needs to be handled by a sub-grid model.
Motivated by \citet{Jana_2018} we assume cosmic ray heating is uniform within each cell in our locally-confined heating model. 
Note this is not to say that the range of cosmic ray heating is the $3$\,cMpc size of the simulation cell, but instead that the overlapping cosmic ray heating regions around different star-forming halos within the cell leads to approximately uniform heating of the IGM within the cell. 
However, variations in heating rate between cells are still present due to the differences in matter overdensity and hence star-forming halo density from cell to cell. 
Furthermore, we assume the heating is partitioned between the ionized and non-ionized phases of each cell in proportion to their masses. 
In Section~\ref{sssec:assumption_breakdown} we discuss the anticipated consequences of these assumptions breaking down, which simulations by \citet{Jana_2018} indicate is likely around large galaxies at redshifts lower than our simulations ($z \sim 4$).

Short-ranged cosmic ray heating such as that we are attempting to describe via our locally-confined heating model would occur if the high-redshift magnetic field in the IGM were strong. 
These strong magnetic fields could either be primordial in nature or originating from weak seed fields amplified by structure formation~\citep{Sur_2010} and then expelled into the IGM via SNRs or galactic outflows~\citep{Samui_2018}. 
Cosmic rays may even somewhat self-confine via the generation of magnetic fields through instabilities or the Biermann battery~\citep{Yokoyama_2022}, though the former process is potentially inefficient in the neutral IGM due to magnetosonic wave damping~\citep{Leite_2017}.

With the two propagation mechanisms (the locally-confined and free-streaming models) bracketing the range of physically plausible distributions for cosmic ray heating, we can now compute the cosmic ray number density in a given cell of our simulation. We combine  the injection rate, equation~\eqref{eq:injected_spectrum_model}, and the chosen propagation window functions
\begin{equation}~\label{eqn:cosmic_ray_field}
\begin{split}
    \frac{d^2N(\mathbfit{x},\ke{}, z)}{dV d\ke{}}  =& \frac{\eta_{\rm cr} }{\mathcal{N}(\alpha_{\rm cr})} \int_{z}^{\infty} dz_{\rm 0} \bigg[- \Big(\kezero{}(z_{\rm 0}, \ke{})\Big)^{\alpha_{\rm cr}}  \left.\frac{dt}{dz}\right|_{\rm z_{\rm 0}} \times \\
    &\Big({\rm SFRD}_{\rm cr}[\mathbfit{x}, z_{\rm 0}] \ast P[\mathbfit{x}, z; \kezero{}(z_{\rm 0}, \ke{}), z_{\rm 0}] \Big) \bigg],
\end{split}
\end{equation}
where $\ast$ is the spatial convolution, and $\kezero{}(z_{\rm 0}, \ke{})$ is the initial kinetic energy computed for a cosmic ray emitted at redshift $z_{\rm 0}$ which has kinetic energy $\ke{}$ at redshift $z$. The function $\kezero{}(z_{\rm 0}, \ke{})$ being defined implicitly as the solution to equation~\ref{eqn:cr_energy_ode}.

\subsubsection{Heating and Ionization Rate}

Finally, we need to convert $d^2N(\ke{}, z)/dV d\ke{}$ into cosmic ray heating and ionization rates. By using the heating fractions $f_{\rm heat}$ defined in subsection~\ref{ssec:cr_heating} we can compute the cosmic ray heating rate per baryon as
\begin{equation} \label{eqn:cr_heating_rate}
    \epsilon_{\rm cr} = \frac{f_{\rm heat}(x_{\rm e})}{n_{\rm b}} \int_0^{\kemax{}} d\ke{} \left[ - \left.\frac{d\ke{}}{dt}\right|_{\rm E\&I}  \frac{d^2N}{dV d\ke{}}\right],
\end{equation}
where $-\left.d\ke{}/dt\right|_{\rm E\&I}$ is the rate of kinetic energy loss by a cosmic ray proton to excitation and ionization interactions. This heating rate can then simply be added to the sum of heating rates in equation~\eqref{eqn:Tk_ode} to model cosmic ray heating in our simulations, with  the corresponding ionization rate from equation~\eqref{eqn:cr_ioniz_rate} used in equation~\eqref{eqn:xe_ode}. To properly account for the dependence of heating on spatial variation of $x_{e}$ (rather than using a globally averaged or fixed value like previous works), we interpolate between the tabulated values of $f_{\rm heat}$ for each cell of our simulation individually.

With our full model established, we can now infer the energy range of cosmic rays that dominate heating. For an individual cosmic ray proton, we consider the fraction of its initial kinetic energy that contributes to the IGM heating by $z = 6$ (shown in Fig.~\ref{fig:fraction_heating}) assuming the particles travel through the IGM of mean density and ionization fraction. As has been found in previous studies~\citep{Sazonov_2015}, the resulting fraction of energy that ends up as IGM heat is strongly dependent on the initial kinetic energy of the cosmic ray proton. Protons with $\kezero{} < 10$\,MeV deposit at least 10 per cent of their initial energy, with this contribution rising to above 30 per cent at $z < 10$ when the free electron fraction increases in the course of reionization. Above $\kezero{} = 10$\,MeV there is an initial gradual decrease in the energy fraction; however, once the cosmic rays have high enough $\kezero{}$ to not be fully absorbed by $z = 6$ the fraction rapidly falls below $1$ per cent. At the highest energy $\kezero{} > 1000$\,MeV less than $0.1$ per cent of a cosmic ray initial kinetic energy contributes to heat. If indeed the cosmic ray injected spectrum has an exponent around $-2$, as is predicted by theory, these results reaffirm the conclusions of \citet{Sazonov_2015} and \citet{Leite_2017} that cosmic ray heating is dominated by lower energy particles with $< 30$\,MeV. However, we also find that cosmic rays with much higher energies up to $200$\,MeV have a non-negligible contribution, heating up the IGM on large scales out to $\sim 100$\,cMpc away from their sources.

\subsubsection{Summary of our cosmic ray model}

For clarity let us summarise our final model for cosmic ray heating. In our framework, cosmic rays are emitted from star-forming halos below a mass threshold $M_{\rm cr}^{\rm max}$ at a rate proportional to either the Pop II, Pop III, or the total star formation rate in that halo, as specified by the user. Particles are injected into the IGM with an efficiency of $\eta_{\rm cr}$ (energy injection per star formation rate), with the injected cosmic ray spectrum taking the form of a power-law in kinetic energy with exponent $\alpha_{\rm cr}$ and lower/upper cutoff of $\kemin{}$/$\kemax{}$. While in the IGM, cosmic rays lose energy to Hubble cooling and excitation and ionization interactions, as described by equation~\eqref{eqn:cr_energy_ode}. Through these processes they heat the IGM at a rate described via equation~\eqref{eqn:cr_heating_rate}, propagating outward via one of three mechanisms: globally uniform, free-streaming, or locally-confined. This propagation is modelled via equation~\eqref{eqn:cosmic_ray_field} using the relevant transfer function: a uniform distribution, a spherical shell (equation~\ref{eqn:cosmic_ray_eom}), and 1 in the origin cell with 0 elsewhere respectively. Cosmic ray heating rate is then integrated into our simulations by extending the sum in equation~\eqref{eqn:Tk_ode}.
The model is flexible by construction with free parameters describing the cosmic ray emitting population ($M_{\rm cr}^{\rm max}$, $\eta_{\rm cr}$, $\alpha_{\rm cr}$, $\kemin{}$, $\kemax{}$) and the propagation mechanisms (globally uniform, free-streaming, locally-confined).

As discussed in Section~\ref{sssec:cr_sources} the efficiency of cosmic ray injection into the IGM, $\eta_{\rm cr}$, is dependant on the fraction of stars that underwent supernovae, supernova kinetic energy yields, the conversion rate of this energy into cosmic rays, and the escape fraction of cosmic rays into the IGM. The first two of these are highly sensitive to the initial mass function of the stellar population of interest. Only higher mass stars undergo supernovae, and the type and hence kinetic energy yield of the supernovae is in turn mass dependent. 
In this study, we treat $\eta_{\rm cr}$ as a free parameter and decouple it from our modelling of the Pop~III initial mass function due to the large uncertainty in the other components of $\eta_{\rm cr}$. Future works that aim to constrain early universe models with 21-cm signal data, could model the link between IMF and $\eta_{\rm cr}$ using appropriate conditional priors.

 While a correspondence cannot always be made between our model parameters and those of other works when it can be it is instructive to consider the parameters used by previous studies. Here for ease of comparison, we attempt to convert the parameters used in various studies to the parameters used in our work, with the values listed in Table.~\ref{tab:prev_study_parameters}. Several of the previous studies consider one supernova of a given yield occurring per halo, which is then converted into cosmic ray energy injected into the IGM with some efficiency parameter or a combination thereof.  To convert such values to $\eta_{\rm cr}$  we have assumed that this supernova is occurring in a $\sim 10^6$\,\solm{} halo, with a baryonic collapse fraction of $0.1$ and a star formation efficiency $f_{\rm *} = 0.01$. Another difference in modelling is the usage of either kinetic energy or momentum power laws. When previous studies have used momentum power laws, we convert their spectral exponents to equivalent kinetic energy power law exponents by equating the proportion of cosmic ray energy that ends up in cosmic rays with $\ke{} < 30$\,MeV, since we anticipate these cosmic rays to dominate heating. The parameter values listed in Table.~\ref{tab:prev_study_parameters} show that the largest differences between previous works are in the values of $\eta_{\rm cr}$ which varies between $1 \times 10^{47}$ and $5 \times 10^{49}$\,erg\,\isolm{}. The discrepancy arises due to the uncertainty in the initial mass function of high redshift stellar populations (and thus type/frequency of supernovae), the efficiency of conversion of supernovae kinetic energy to cosmic rays, and the escape fraction of these cosmic rays into the IGM.

\begin{table*}
 \caption{Parameters used by previous comic ray heating studies converted into our notation, in order: the emitting stellar population, the maximum halos mass cosmic rays can escape from $M_{\rm cr}^{\rm max}$, the efficiency of cosmic ray emission $\eta_{\rm cr}$, the exponent of the cosmic ray spectrum $\alpha_{\rm cr}$, and the minimum/maximum cutoff of the cosmic ray spectrum $\kemin{}$/$\kemax{}$. The propagation mechanism of cosmic rays is not included as all the previous studies assumed uniform heating, with the exception of \citet{Jana_2018} who considered individual halos. Values marked with a red asterisk are approximate equivalents. If there is no close equivalent value, or no value is given in the corresponding paper, we leave that table entry unpopulated (dash). Two separate sets of parameters are given for \citet{Bera_2022} as they model Pop II and Pop III stars separately. For ease of comparison, we also list the ranges of values used in our subsequent investigation broken down by section.}
 \label{tab:prev_study_parameters}
 \begin{tabular}{lcccccc}
  \hline
  Study & Emitting Pop. & $M_{\rm cr}^{\rm max}$ [\solm{}] & $\eta_{\rm cr}$ [erg\,\isolm{}] & $\alpha_{\rm cr}$ & $\kemin{}$ [MeV] &  $\kemax{}$ [MeV]\\
  \hline
  \citet{Sazonov_2015} & III & $10^7$ & $1 \times 10^{47}$ to $5 \times 10^{49}$\rast{} & $-2$ & $10^{-3}$ & $10^{8}$ \\
  \citet{Leite_2017} & II & $\infty$ & $1 \times 10^{48}$ & $-1.74$ to $-1.94$\rast{} & $10^{-2}$ & $10^9$  \\
  \citet{Jana_2018} & - & - & $6 \times 10^{47}$ & $-1.97$\rast{} & $0.1$ and $1$ & - \\
  \citet{Jana_2019} & III & - & $1.5 \times 10^{49}$\rast{} & - & - & - \\
  \citet{Bera_2022} (Pop II) & II & $\infty$ & $3 \times 10^{48}$ & - & - & - \\
  \citet{Bera_2022} (Pop III) & III & $\infty$ & $4 \times 10^{48}$\rast{} & $-1.84$\rast{} & $10^{-2}$ & - \\
   \hline 
   Section~\ref{ssec:clustering} & II + III  & $\infty$  &  $1 \times 10^{48}$ to $1.39 \times 10^{48}$&  $-2$  &  $10^{-2}$  & $10^9$  \\
   Section~\ref{sssec:global} & II + III  & $\infty$  &  $1 \times 10^{48}$&  $-2$  &  $10^{-2}$  & $10^9$  \\
   Section~\ref{ssec:heating_comparison} & II + III  & $\infty$  &  $1 \times 10^{47}$ to $1 \times 10^{49}$&  $-2$  &  $10^{-2}$  & $10^9$  \\
   Section~\ref{ssec:diffusive} & II + III  & $\infty$  &  $1 \times 10^{48}$ & $-1.7$ to $-2$  &  $10^{-2}$  & $10^9$  \\
   Section~\ref{ssec:direct} & III  & $10^6$ to $10^8$  &  $3 \times 10^{48}$ & $-2$  &  $10^{-3}$  & $10^8$  \\
   \hline
 \end{tabular}
\end{table*}

A further potential impact of cosmic rays at high redshift is positive feedback on star formation via enhancing the ${\rm H}_{\rm 2}$ concentration \citep{Stacy_2007, Jasche_2007}, however, we do not include this effect in our simulations due to these results being refuted by \citet{Hummel_2016}. We also do not include the negative feedback from cosmic rays heating gas near the halos virial radius which might suppress the rate of gas accretion onto the halo \citep{Lacki_2015,Jana_2018}, as it is only efficient at low redshifts ($z \leq 4$).

Finally and importantly, we do not include excess radio backgrounds from cosmic rays. Although  \citet{Jana_2019} demonstrated that cosmic ray electrons in Pop III SNRs can produce an excess radio background via synchrotron emission, their analysis assumed an ambient magnetic field of $0.32$\,$\mu$G at $z = 17$ (equivalent to $\sim 1$\,nG comoving). As we discussed earlier, such a strong magnetic field might be too extreme as it is at the upper bound of the extremely broad (11 orders of magnitude) range of the experimentally allowed primordial magnetic field values. The excess radio background found by \citet{Jana_2019} scales as the magnetic field to the $8/5$th power, and so an order of magnitude lower magnetic field would reduce the extra radio background relative to the CMB from 300 per cent to only 8 per cent. If we assume a comoving magnetic field of $\lesssim 0.01$\,nG consistent with our treatment of Alfv\'en wave heating, the excess radio background from cosmic rays should be less than a per cent of the CMB temperature at the redshifts of interest, and thus will have negligible impact on the 21-cm signal~\citep{Sikder_2023}.

\section{Results}
\label{sec:results}

We now describe our findings regarding the impacts of cosmic ray heating on 21-cm observables, splitting our investigations into four main themes. 
First, in subsection~\ref{ssec:clustering} we explore the implications of the short-range nature of cosmic ray heating on the 21-cm signal and how this could be leveraged to distinguish a cosmic ray heated IGM from an IGM heated by other mechanisms. 
Secondly, we discuss the biases introduced by assuming spatially uniform cosmic ray heating in subsection~\ref{sssec:global}.
Thirdly in subsection~\ref{ssec:heating_comparison}, we compare the efficiency of cosmic ray heating to X-ray heating. 
Finally, we contrast the impact of cosmic ray heating with other high-redshift astrophysical processes in subsection~\ref{ssec:astrophysics}.

\subsection{Impacts of cosmic ray heating clustering} \label{ssec:clustering}

Previously (subsection~\ref{ssec:cr_heating_summary}) we arrived at the same conclusions as ~\citet{Sazonov_2015, Leite_2017, Jana_2018} that cosmic ray heating is dominated by sub-relativistic protons and, thus, is likely clustered around overdense regions. This is in contrast to heating by X-ray sources~\citep{Furlanetto_2006, Fialkov_2014b, Pacucci_2014} which affects the gas temperature on large cosmological scales of several hundred cMpc. To isolate the characteristic patterns of cosmic ray heating in the 21-cm signal from the effects of other processes we compare models with similar IGM thermal histories. 

Our main case ({\it reference model}) is similar to the model considered by \citet{Leite_2017}, with cosmic rays emitted from all star-forming halos regardless of population or mass (with parameters $\eta_{\rm cr} = 10^{48}$\,erg\,\isolm{}, $\kemin{} = 10^{-2}$\,MeV, $\kemax{} = 10^{9}$\,MeV, $\alpha_{\rm cr} = -2$) and assuming a free-streaming mode of cosmic ray propagation. We contrast this model with locally-confined cosmic ray heating, X-ray heating with a soft spectral energy distribution (SED), and X-ray heating with a hard SED. In our simulations, a soft X-ray SED is modelled as a truncated power-law with exponent $-1.5$ and lower cutoff of $0.2$\,keV corresponding to a mean free path of $\sim 1.3$\,cMpc at $z = 15$~\citep{Furlanetto_2006}. The majority of X-ray photons in this case have mean free paths of several cMpc and, thus, inject energy relatively close to the sources (although are still longer-range compared to cosmic rays). The hard SED is modelled using a truncated power-law with exponent $-1$ and lower-cutoff of $3$\,keV, corresponding to a mean free path of $\sim 4300$\,cMpc at $z = 15$. In this case, most photons travel large cosmic distances, lose most of their energy to cosmic expansion and are hardly absorbed by the IGM. Therefore, this model is expected to yield relatively uniform and weak heating.

\begin{figure}
	\includegraphics{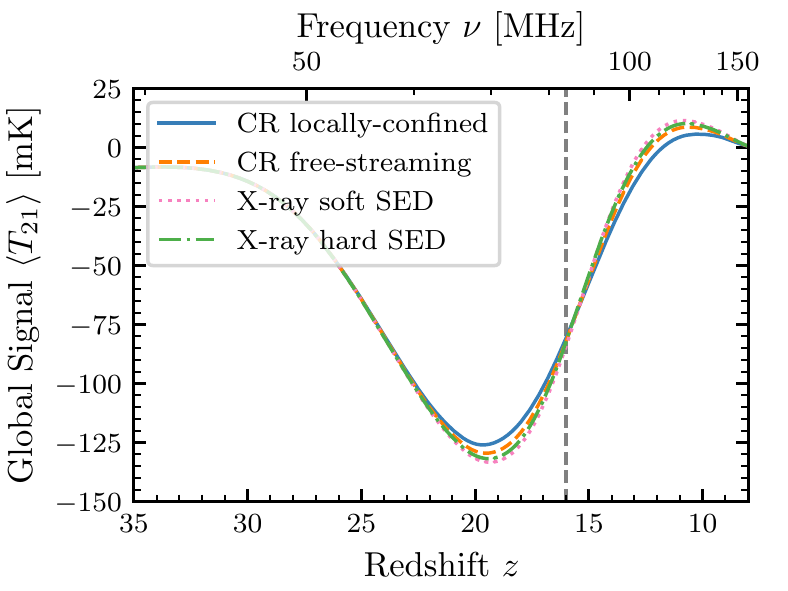}
    \caption{Matched 21-cm global signals with different heating mechanisms. The efficiency of cosmic ray/X-ray emission was tuned for the models to minimize the root-mean-square error between their predicted global signals and that of the reference free-streaming cosmic ray heating model with  $\eta_{\rm cr} = 10^{48}$\,erg\,\isolm{}. This procedure gave $\eta_{\rm cr} = 1.39 \times 10^{48}$\,erg\,\isolm{} for the locally-confined cosmic ray heating model, $f_{\rm X} = 0.332$ for the soft SED X-ray heating model, and $f_{\rm X} = 60.0$ for the hard SED X-ray heating model. All other astrophysical parameters listed in Table \ref{tab:sim_default_parameters} and cosmological parameters, including the cosmological initial conditions, are kept the same between the simulations. The four models show good agreement at $z > 22$ as heating of the IGM has yet to become significant, as well as in the heating arm of the global signal around $z \sim 15$. Differences are still seen between models at the global signal minimum and emission maximum. A vertical grey dashed line is shown at $z = 16$, illustrating the redshift at which we compare tomographic 21-cm maps, see Fig.~\ref{fig:tomographic_cubes}.}
    \label{fig:global_signal_normalization}
\end{figure}

For ease of comparison, we calibrate heating efficiency parameters (either $\eta_{\rm cr}$ or $f_{\rm X}$\footnote{All other astrophysical parameters (listed in Table \ref{tab:sim_default_parameters})  and cosmological parameters, including the cosmological initial conditions, are kept the same between the simulations.}) to achieve similar thermal histories between the models.  We achieve this by minimizing the root-mean-square difference between the reference global 21-cm  signal\footnote{We also considered minimizing the difference between average kinetic temperatures and came to the same conclusions as we do here for matching global 21-cm signals.} and the global signals of each other model over the redshift range $z = 10-25$. Through this procedure, we find signals with minimal differences as shown in  Fig.~\ref{fig:global_signal_normalization}. We find that the reference model with $\eta_{\rm cr} = 10^{48}$\,erg\,\isolm{} produces a similar global signal and a similar thermal history as the locally-confined cosmic ray heating model with $\eta_{\rm cr} = 1.39 \times 10^{48}$\,erg\,\isolm{}, soft X-rays with $f_{\rm X} = 0.332$ and hard X-rays with  $f_{\rm X} = 60.0$. For comparison, we note that $f_X = 1$ is the typical value calibrated to the present-day population of X-ray binaries and taking into account the more efficient X-ray emission in metal-poor environments of the high redshift universe \citep{Fragos_2013}.

Closely comparing the matched global signals we see that the high-redshift parts regulated by the onset of star formation and WF coupling  ($z \sim 25-35$) are identical for all the calibrated models. Small differences in the shapes of the signals arise at redshifts affected by heating processes ($z \sim 10-20$, we refer to this part of the signal as {\it the heating arm}) suggesting that differences in the spatial distribution of heating do not fully average out. In the following, to produce the clearest visual comparison of heating signatures, we contrast signals at $z = 16$ when the four models have approximately equal global signals $\langle T_{\rm 21} \rangle = 82$--$85$\,mK.

\subsubsection{Imprints of cosmic ray heating in 21-cm tomography} \label{sssec:tomography}

The 21-cm signal is sensitive to gas temperature and, therefore, the character of the IGM heating is expected to be manifested in the spatial distribution of the brightness temperature. To visualize the differences in heating patterns between cosmic rays and X-rays we begin by considering 21-cm tomography. 

\begin{figure*}
	\includegraphics[width=\textwidth]{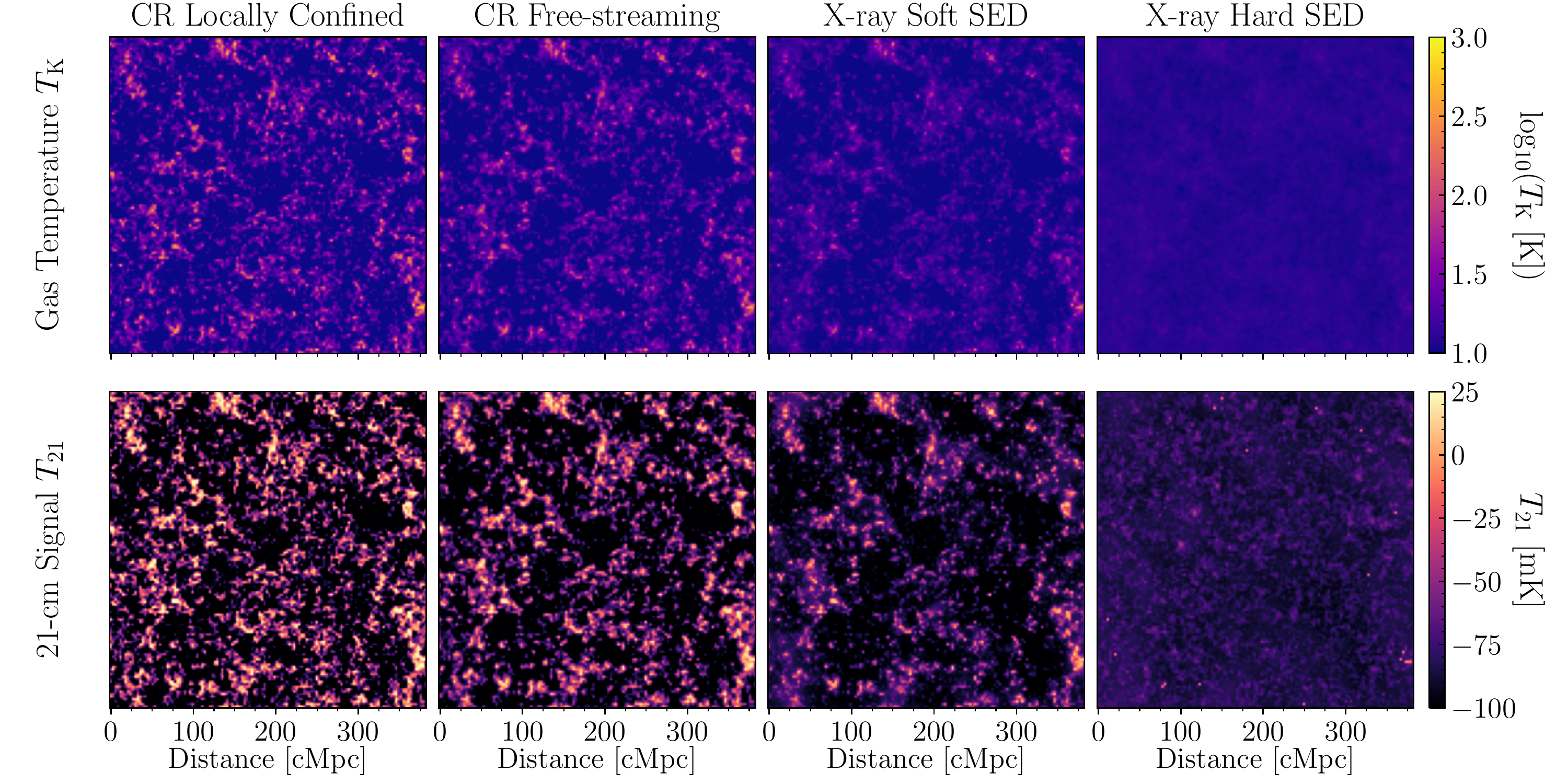}
    \caption{Slices of tomographic maps of the IGM gas temperatures (top) and the 21-cm signals (bottom) for different cosmic ray (CR) and X-ray heating mechanisms shown at $z = 16$ where all four simulations predict approximately the same global signal $\langle T_{\rm 21} \rangle \approx -84$\,mK. From left to right the IGM heating becomes more diffused as the heating carriers become longer-ranged (locally-confined CRs, free-streaming CRs, soft X-rays and hard X-rays). In addition, the mean IGM gas temperature decreases from $20.5$\,K to $16.0$\,K, $13.4$\,K, and finally $12.4$\,K. The change in heating results in a reduced contrast between emitting and absorbing regions in the 21-cm maps.  All four simulations used the same cosmological initial conditions and the same slice of the simulation box is shown in each case.  }
    \label{fig:tomographic_cubes}
\end{figure*}

Fig.~\ref{fig:tomographic_cubes} shows slices of the gas temperature cubes at $z = 16$ produced by the four simulations outlined above, alongside the corresponding slices of the 21-cm signal. As anticipated, the maps show that in cosmic ray heated models, the heating of the IGM is much more clustered compared to the cases with X-ray heating. We see a sharp contrast between the localized high-temperature regions and the vast regions of cooler gas, which is manifested in the 21-cm maps as regions of mild emission signal and deep absorption. The contrast is the sharpest for locally-confined cosmic ray heating as is expected given it has the shortest range among the considered scenarios. As we consider increasingly longer-ranged heating mechanisms (i.e. free-streaming cosmic rays, X-ray heating with soft SED, and, finally, hard X-rays), fluctuations in the gas temperature are reduced, and, consequently, the brightness temperature maps show less contrast. Ultimately, in the case of heating by hard X-rays (the rightmost column), the gas temperature is nearly uniform and there are practically no visible emission regions in the 21-cm maps. This trend is explained by the increasingly larger mean free path of the energy carriers (either cosmic ray protons or X-ray photons). For X-ray heating to appear as clustered as cosmic ray heating, the mean-free path of X-rays would need to be at most $\sim $3\,cMpc which at $z = 16$ (and assuming neutral IGM) corresponds to X-ray energies of $\sim 280$\,eV or lower. However, high-redshift X-ray sources, such as X-ray binaries, are thought to have much harder SEDs peaking at a few keV ~\citep{Fragos_2013, Sartorio_2023}. Therefore, our simulations suggest that, if observed, the character of the heated regions in the 21-cm maps could be used to probe the nature of the dominant heating mechanism, e.g. distinguish between cosmic ray and X-ray heating.

\subsubsection{Signatures of heating clustering in the 21-cm power spectrum} \label{sssec:power}

\begin{figure*}
	\includegraphics[width=\textwidth]{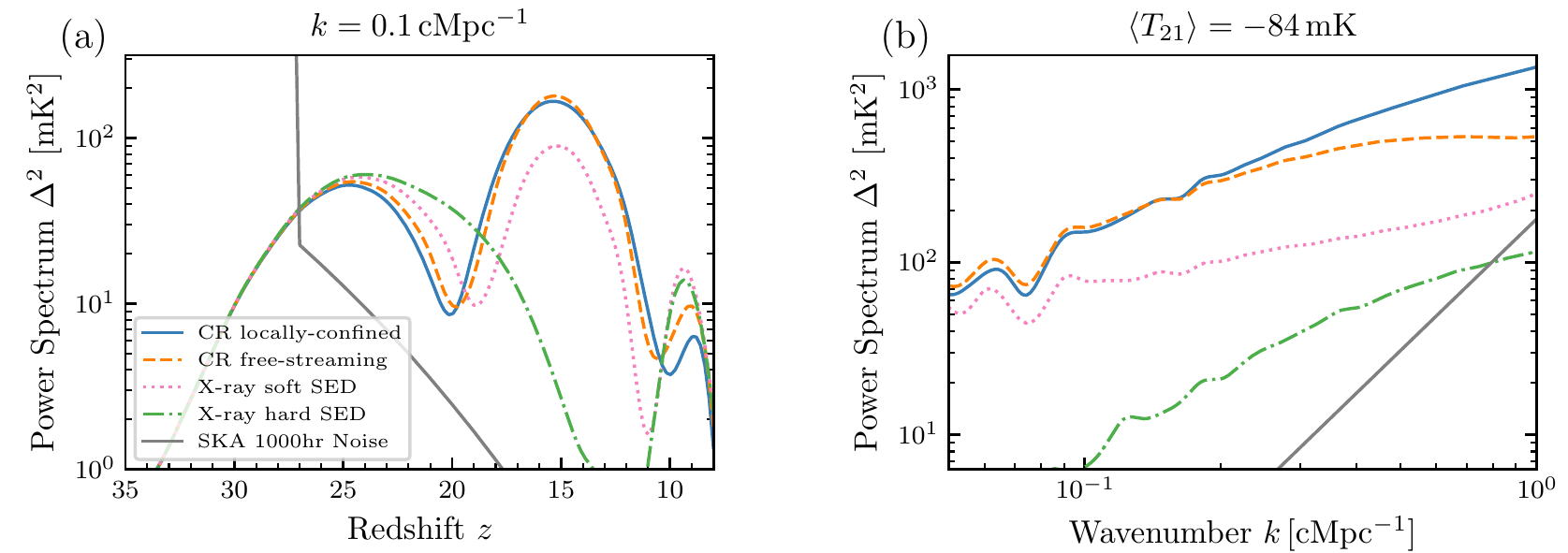}
    \caption{Comparison of the 21-cm power spectrum for the locally-confined cosmic rays, free-streaming cosmic rays, soft and hard X-rays (models are specified in subsection~\ref{sssec:tomography}). Panel (a) shows the redshift evolution of the 21-cm power spectrum at a fixed wavenumber $k = 0.1$\,cMpc$^{-1}$. Panel (b) shows the power spectra versus wavenumber $k$ at the redshift when the corresponding global 21-cm signal equals to $\langle T_{\rm 21} \rangle = -84$\,mK (which occurs at $z\sim16$ for all the considered cases). We find that the small-scale (high $k$) 21-cm signal varies significantly between the models, suggesting that the variation of the 21-cm power spectrum with $k$ could provide a diagnostic tool for the dominant heating mechanism. Also shown in both panels is the thermal-noise estimate for the SKA with 1000\,hrs of observations~\citep[grey curve, ][]{Koopmans_2015}, illustrating the theoretical sensitivity of next-generation 21-cm power spectrum experiments.}
    \label{fig:power_spectrum_diagnostic}
\end{figure*}

The 21-cm power spectrum is a signal targeted by radio interferometers including HERA~\citep{HERA}, LOFAR~\citep{LOFAR}, MWA~\citep{MWA}, NenuFAR~\citep{NenuFAR}, and the future SKA~\citep{Koopmans_2015}.   
Examples of the 21-cm power spectra for the models explored here are shown in Fig.~\ref{fig:power_spectrum_diagnostic}.  We start by considering the redshift evolution of the power spectra at a fixed comoving wavenumber $k = 0.1$\,cMpc$^{-1}$ (panel~a). In general, we expect to see two well-defined peaks in the power spectra vs redshift, one marking the dominance of fluctuations imprinted by the non-uniform WF coupling and the other reflecting the gas temperature fluctuations prevailing at lower redshifts. We find the power spectra produced by the two cosmic ray heating scenarios to be very similar while being dramatically different from the X-ray heating scenarios. As a result of their localized strong heating, the two cosmic ray models show clear narrow peaks due to the WF coupling at $z\sim 25$ and heating at $z\sim 15$. In the soft X-ray SED case heating is longer-range which results in a slight delay in energy injection, a broader WF coupling peak and a factor of $1.8$ lower heating peak. Finally, in the hard X-ray SED case the Lyman-coupling peak is even broader and the heating peak is completely erased owing to the large mean free path of X-ray photons. 

Next, we compare the shape of the signals as a function of wavenumber $k$ at a fixed global signal value. For convenience, we choose the value of $\langle T_{\rm 21} \rangle = -84$\,mK\footnote{ The same comparison was performed for $\langle T_{\rm 21} \rangle = -50$\,mK and $\langle T_{\rm 21} \rangle = -100$\,mK with similar conclusions reached.} (which for all the models happens at $z\sim 16$) to show results for in panel~b.  The differences seen in the shapes of the power spectra for various heating mechanisms are much more striking than what we saw for the corresponding global signals and can be understood in terms of the typical length scale below which the structure is washed out. In the locally-confined cosmic ray heating, this length scale is at the resolution limit of our simulation and so the spherical 21-cm power spectrum traces the matter power spectrum across our entire $k$ range. For cosmic ray free-streaming the heating length scale is larger leading to a suppression of the power spectrum above $k \sim 0.2$\,cMpc$^{-1}$, which is evident from the figure. In the case of X-ray heating, the high energy particles (present even in the soft SED case) affect the shape of the signal at a broad range of scales leading to a suppression of power. The effect is visible even at the largest scales (smallest $k$) considered, though the suppression remains strongest at smaller scales (high $k$). Finally, since for the hard SED X-ray model the heating is nearly uniform, the strong suppression of the power spectrum is observed at all the considered scales. In summary, we find that the strongest differences in the signatures of heating mechanisms are produced at high $k$, with the locally-confined cosmic ray heated model having a $60$ per cent, $420$ per cent, and $1080$ per cent larger power spectrum than the free-streaming cosmic ray, soft X-ray SED, and hard X-ray SED models respectively at $k = 0.5$\,cMpc$^{-1}$.

Our results suggest that a measurement of the 21-cm power spectrum at $k > 0.2$\,cMpc$^{-1}$ during the heating-driven stage ($z \sim 10-20$) could be used to probe the dominant IGM heating mechanism. Stronger signals at high $k$ values would indicate a more clustered heating, allowing us to distinguish X-ray from cosmic ray heating, and (if the environment was cosmic ray heated) the degree of confinement of cosmic ray particles. To illustrate the experimental feasibility of using the 21-cm power spectrum as such a diagnostic tool we have included 1000~hr SKA thermal noise sensitivities~\citep{Koopmans_2015} in Fig.~\ref{fig:power_spectrum_diagnostic}, which yield signal-to-noise-ratios $> 9$ at $k = 0.5$\,cMpc$^{-1}$ between each pair of models. 
In the heating-driven era of the 21-cm signal the suppression of the 21-cm power spectrum above a length-scale set by the dominant heating mechanism is a generic effect, previously proposed by \citet{Fialkov_2014b} as a way to probe the SED of X-ray sources. 
Thus the potential to distinguish the nature of high-redshift heating mechanisms based on the shape of the high $k$ 21-cm power spectrum should be robust to astrophysical uncertainties, though the exact signal-to-noise ratios will depend on the astrophysical scenario.
Additionally, this analysis does not consider experimental systematics, or any potential degeneracies with new physics such as properties of dark matter~\citep{Sitwell_2014, Barkana_2018, Munoz_2018, Fraser_2018, Fialkov_2018, Liu_2019, Munoz_2020, Jones_2021, Hibbard_2022, Barkana_2022}, all of which could weaken the ability for the high $k$ power spectrum to pin down the dominant IGM heating mechanism. Given these uncertainties, further work is required to reliably evaluate the statistical significance with which the nature of high-redshift heating mechanisms can be determined from the projected 21-cm power spectrum measurements of upcoming experiments.

\subsection{Uniform heating assumption and the global signal} \label{sssec:global}

When attempting to match  thermal histories between simulations~(Fig.~\ref{fig:global_signal_normalization}), we found that an exact correspondence between the produced global signals could not be achieved in models with locally-confined and free-streaming cosmic ray heating. The inevitable discrepancy suggests that the spatial distribution of IGM heating is reflected in the global 21-cm signal due to $\langle T_{\rm 21} \rangle$ being a non-linear tracer of $T_{\rm K}$. Given that previous studies assumed uniform cosmic ray heating, rather than the theoretically expected highly clustered behaviour, such an assumption may have biased the global signal predictions in these works. 

\begin{figure}
	\includegraphics{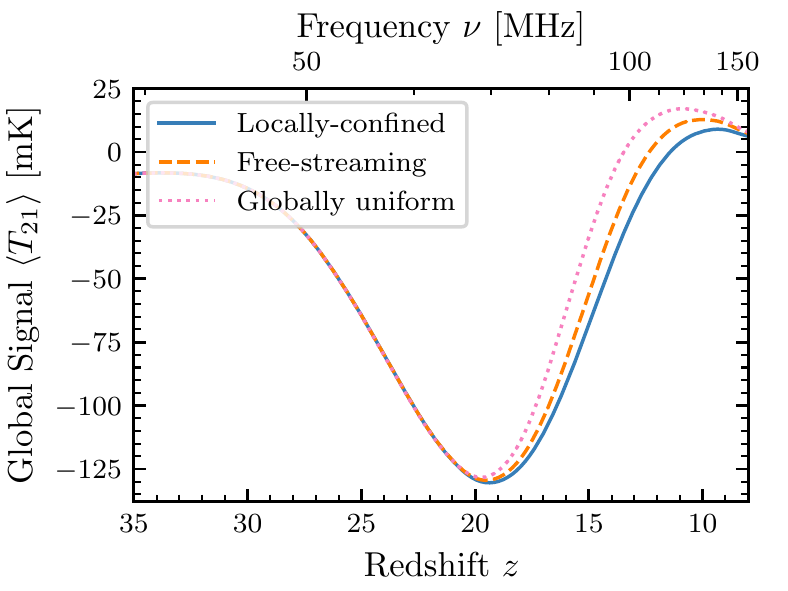}
    \caption{Global 21-cm signal for different cosmic ray propagation models. All three models have the same cosmic ray emission efficiency and astrophysical parameters, and thus the energy in cosmic rays reaching the IGM is the same between the models. At redshifts affected by heating ($z \sim 12-20$), the signal in the globally uniform cosmic ray heating model evolves faster and reaches a higher emission peak than the other two propagation models. Similar but smaller differences can also be seen between the free-streaming and locally-confined models. These differences suggest that clustered heating is less efficient than diffused heating at driving the 21-cm global signal into emission.}
    \label{fig:global_signal_propogation.pdf}
\end{figure}

Here we investigate the degree to which the global signal is affected by the locality of heating. Unlike in the previous subsection, here we do not attempt to match thermal histories between models, instead, we keep the cosmic ray emission efficiency parameter \mbox{$\eta_{\rm cr} = 10^{48}$\,erg\,\isolm{}} constant so that the energy injected into cosmic rays that reach the IGM is the same between simulations. This allows us to isolate the effect of the heating distribution by only varying the cosmic ray propagation mode between locally-confined, free-streaming and globally uniform heating. All other parameters are kept the same as for the reference model used in the previous subsection (and listed in Table \ref{tab:sim_default_parameters}). The 21-cm global signals predicted for the three different cosmic ray propagation modes are shown in Fig.~\ref{fig:global_signal_propogation.pdf}.

As expected, the differences in the predicted global signals with various heating mechanisms are manifested at redshifts affected by heating ($z \lesssim 20$), while the signals are identical at higher redshifts dominated by the WF coupling. In the case of uniform heating, the global signal evolution happens faster, preceding the other two models by $\Delta z \sim 1$. The uniform heating model also features a higher and earlier emission maximum of $\langle T_{\rm 21} \rangle = 17.9$\,mK, at $z = 11$ compared to $\langle T_{\rm 21} \rangle = 12.8$\,mK, at $z = 10$ and $\langle T_{\rm 21} \rangle = 8.7$\,mK, at $z = 10$ in the free-streaming and locally-confined cases respectively. While the effect is smaller we also observe similar differences between the free-streaming and locally-confined propagation modes.

Even though the energy of cosmic rays reaching the IGM is the same in all cases, our results suggest that clustered heating is less efficient at increasing the average 21-cm signal than uniform heating. This is contrary to what would be expected from the mean IGM kinetic temperature, which is found to be highest between $z = 13$ and $20$ for the locally-confined model, due to the correlations between star formation and $x_{\rm e}$ which increase the average efficiency with which cosmic ray energy is deposited into the IGM as heat ($f_{\rm heat}$). The non-linear relationship between $\langle T_{\rm 21} \rangle$ and $T_{\rm K}$ explains the fact that uniform heating is the most efficient at increasing $\langle T_{\rm 21} \rangle$. When the WF coupling is efficient ($x_{\alpha} \gg 1$) but before reionization becomes significant (i.e. $x_{\rm HI} \approx 1$), the 21-cm signal scales as $T_{\rm 21} \sim 1 - T_{\gamma}/T_{\rm K}$. From this concave functional dependence, we see that $T_{\rm 21}$ saturates at a positive value for $T_{\rm K} \gg T_{\gamma}$ but can take large negative values for $T_{\rm K} \ll T_{\gamma}$. Consequently, when a volume average of $T_{\rm 21}$ is taken the weighting favours low $T_{\rm K}$. As a result, owing to the concentration of heating into a small volume (that then saturates in 21-cm emission), models with clustered IGM heating are less efficient at raising the mean $T_{\rm 21}$ signal than the models in which heating is distributed evenly across the entire IGM. This explains our findings that the shorter range cosmic ray heating we consider the less effective the heating appears to be at raising $\langle T_{\rm 21} \rangle$.

The assumption of uniform cosmic ray heating has thus somewhat biased the predictions of previous studies, with cosmic ray heating appearing too effective at increasing the 21-cm global signal. It should be noted, however, that this effect is moderate with the largest discrepancy being $34.6$\,mK at $z = 15$, between locally-confined and globally uniform heating mechanisms, comparable to the $25$\,mK EDGES residuals~\citep{EDGES}, though at a potentially measurable level for future experiments~\cite[e.g. the $5$\,mK sensitivity level projected for 2500 hours of REACH observations, ][]{REACH}. Hence, this bias while present is not anticipated to greatly impact the conclusions of previous studies.

\subsection{Comparison of heating efficiencies} \label{ssec:heating_comparison}

\begin{figure}
	\includegraphics{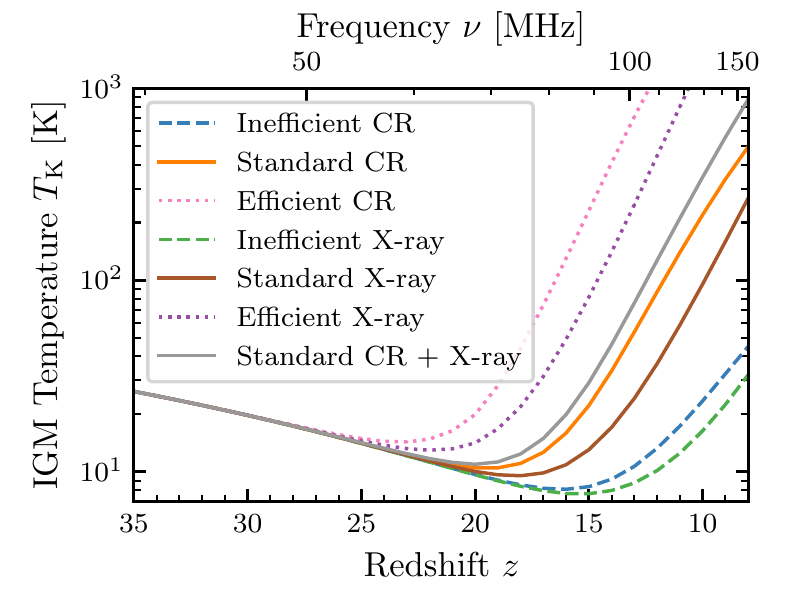}
    \caption{Evolution of the volume-averaged IGM temperature outside of ionized bubbles for different heating mechanisms. Shown are the kinetic temperature for models with inefficient, standard, and efficient cosmic ray (CR) heating, corresponding to $\eta_{\rm cr} = 10^{47}$\,erg\,\isolm{}, $10^{48}$\,erg\,\isolm{}, and $10^{49}$\,erg\,\isolm{} respectively; inefficient, standard, and efficient X-ray heating, corresponding to $f_{\rm X} = 0.1$, $1$, and $10$; as well as a model with both standard efficiency cosmic ray heating ($\eta_{\rm cr} = 10^{48}$\,erg\,\isolm{}) and X-ray heating ($f_{\rm X} = 1$). We find that at standard efficiencies considered in the literature, cosmic ray heating heats up the neutral IGM more rapidly than X-rays. However, as both mechanisms at high redshifts are very uncertain, we advocate for the inclusion of both. }
    \label{fig:heating_differences}
\end{figure}

So far we have considered simulations with either cosmic ray heating or X-ray heating active. However, in reality, both mechanisms will take place simultaneously. Here we contrast the two processes to gain insight as to under which circumstances each process would have a dominant contribution to thermal history. We consider three simulations modelling cosmic ray heating only, with inefficient ($\eta_{\rm cr} = 10^{47}$\,erg\,\isolm{}), standard ($\eta_{\rm cr} = 10^{48}$\,erg\,\isolm{}), or efficient ($\eta_{\rm cr} = 10^{49}$\,erg\,\isolm{}) cosmic ray emission, and a further three simulations modelling X-ray heating only, with inefficient ($f_{\rm X} = 0.1$), standard ($f_{\rm X} = 1$), or efficient ($f_{\rm X} = 10$) X-ray emission which use the X-ray SED calculated by \citet{Fragos_2013} for early universe X-ray binaries. One should keep in mind that the selected efficiencies are illustrative and both heating processes are subject to orders of magnitude uncertainties. For comparison, we also consider a seventh simulation with both standard efficiency cosmic ray heating and standard X-ray heating modelled. All other parameters of these simulations are the same as the reference simulation from subsection~\ref{sssec:tomography}. The predicted evolution of the volume-averaged IGM temperature outside of ionized bubbles for these simulations is shown in Fig.~\ref{fig:heating_differences}.

We find that for the range of typical efficiencies considered in the literature, cosmic rays are roughly as efficient as X-rays in heating up the neutral IGM at $z < 20$. 
Comparing the cosmic ray and X-ray standard efficiencies that we adopt here, cosmic rays are found to be slightly more efficient than X-rays at raising the temperature of the IGM.  Furthermore, we see the temperature for the simulation with both cosmic ray heating and X-ray heating modelled is noticeably above that of the simulations with the standard efficiency heating mechanisms modelled individually. This fact is a manifestation of the comparable heating rates and shows that for some plausible efficiency parameter values, neither heating mechanism dominates and modelling of both is required to produce an accurate thermal history. This comparison indicates the potential importance of cosmic ray heating to the modelling of the 21-cm signal.

\subsection{Sub-grid clustering of cosmic ray heating} \label{sssec:assumption_breakdown}

In the above, motivated by the findings of \citet{Jana_2018}, we have assumed that cosmic ray heating is uniform within simulation cells.
However, this assumption is not expected to hold around massive galaxies at $z \sim 4$~\citep{Jana_2018} and may break down at $z > 20$ when there is a greater distance between star-forming halos.
If the assumption were not valid, we would be artificially smoothing cosmic ray heating, and thus the 21-cm signal, on the scale of the resolution of our simulations. 
It is thus necessary to consider the impact on our conclusions of clustering of cosmic ray heating on scales smaller than our resolution.

There are two main impacts of sub-grid heating clustering to consider the inhomogeneity of the temperature of the neutral IGM within a cell, and the proportion of cosmic ray heat deposited into the ionized IGM. 
In section~\ref{sssec:global}, we previously discussed how increased inhomogeneity in the temperature of the IGM reduces the effectiveness of a heating mechanism at raising the 21-cm global signal, thereby delaying the rise out of the global signal absorption trough~(see Fig.~\ref{fig:global_signal_propogation.pdf}). 
The theoretical explanation for this phenomenon ultimately stems from the non-linear relationship between $T_{\rm 21}$ and $T_{\rm K}$, and so will also apply to these subgrid scales. 
Hence sub-grid clustering of heating would lead to a slower rise of the 21-cm global signal post absorption minimum compared to our current models with the same $\eta_{\rm cr}$.
Consequently, a higher $\eta_{\rm cr}$ would be needed to match cosmic ray heating scenarios to X-ray heating scenarios as we did in Section~\ref{sssec:tomography}.
Additionally, we would expect the trends seen when comparing 21-cm tomographic maps in Section~\ref{sssec:tomography} to continue, with more clustered heating leading to greater contrast in the 21-cm maps around radiative sources and larger voids with nearly uniform 21-cm signal.
As a result, the wavenumber above which the 21-cm power spectrum is suppressed would now be set by the sub-grid length scale at which cosmic ray heating is clustered, leading the 21-cm power spectrum to trace the matter power spectrum to even higher wavenumbers than were seen in our locally-confined case.

If heating is clustered around sources such that the heating of the IGM between them is not approximately uniform, then we would also anticipate a greater proportion of the cosmic ray energy to be deposited in the ionized regions close to the sources.
This is in contrast to our current assumption that heating in a cell is split between the ionized and non-ionized IGM in proportion to their masses. 
Since fully-ionized regions have zero 21-cm signal this would further diminish the efficiency of cosmic ray heating at raising the 21-cm signal from its minimum.
However, providing that cosmic rays can still escape the ionized regions surrounding galaxies the length scale at which the 21-cm power spectrum begins to deviate from tracing the matter power spectrum will still be set by the heating length scale.
As a result, as long as there is still a well-defined heating-driven era of the 21-cm signal the 21-cm power spectrum should provide a diagnostic tool for probing the heating mechanism through its characteristic length scale.

Hence, overall we find that should cosmic ray heating not be uniform within cells we would expect a greater contrast in 21-cm signal tomographic maps near sources and larger unheated voids with nearly uniform 21-cm signal away from sources. In addition, we anticipate there would be a slower increase in the 21-cm global signal, and the 21-cm power spectrum would deviate from the matter power spectrum at higher wavenumbers. 
These three changes would serve to enhance our conclusions that X-ray heating and cosmic ray heating can be distinguished using the small-scale 21-cm power spectrum, and that assuming globally uniform cosmic ray heating makes cosmic ray heating erroneously efficient at raising the 21-cm signal.

\subsection{Sensitivity to astrophysics} 
\label{ssec:astrophysics}

In the previous subsections, we considered models with a fixed set of astrophysical parameters listed in Table \ref{tab:sim_default_parameters}, and only varied the type of the IGM heating, cosmic ray propagation model, and emission efficiencies. However, both the cosmic ray heating mechanism and the astrophysical parameters of the early universe are highly uncertain. In this subsection, we consider how the strength and behaviour of cosmic ray heating vary when varying other astrophysical parameters of the model. Throughout this subsection, all simulations have cosmic ray heating, \lya{} heating and CMB heating enabled, X-ray heating disabled, and use the free-streaming propagation model for cosmic rays.

\subsubsection{Diffusive escape}~\label{ssec:diffusive}

\begin{figure*}
	\includegraphics[width=\textwidth]{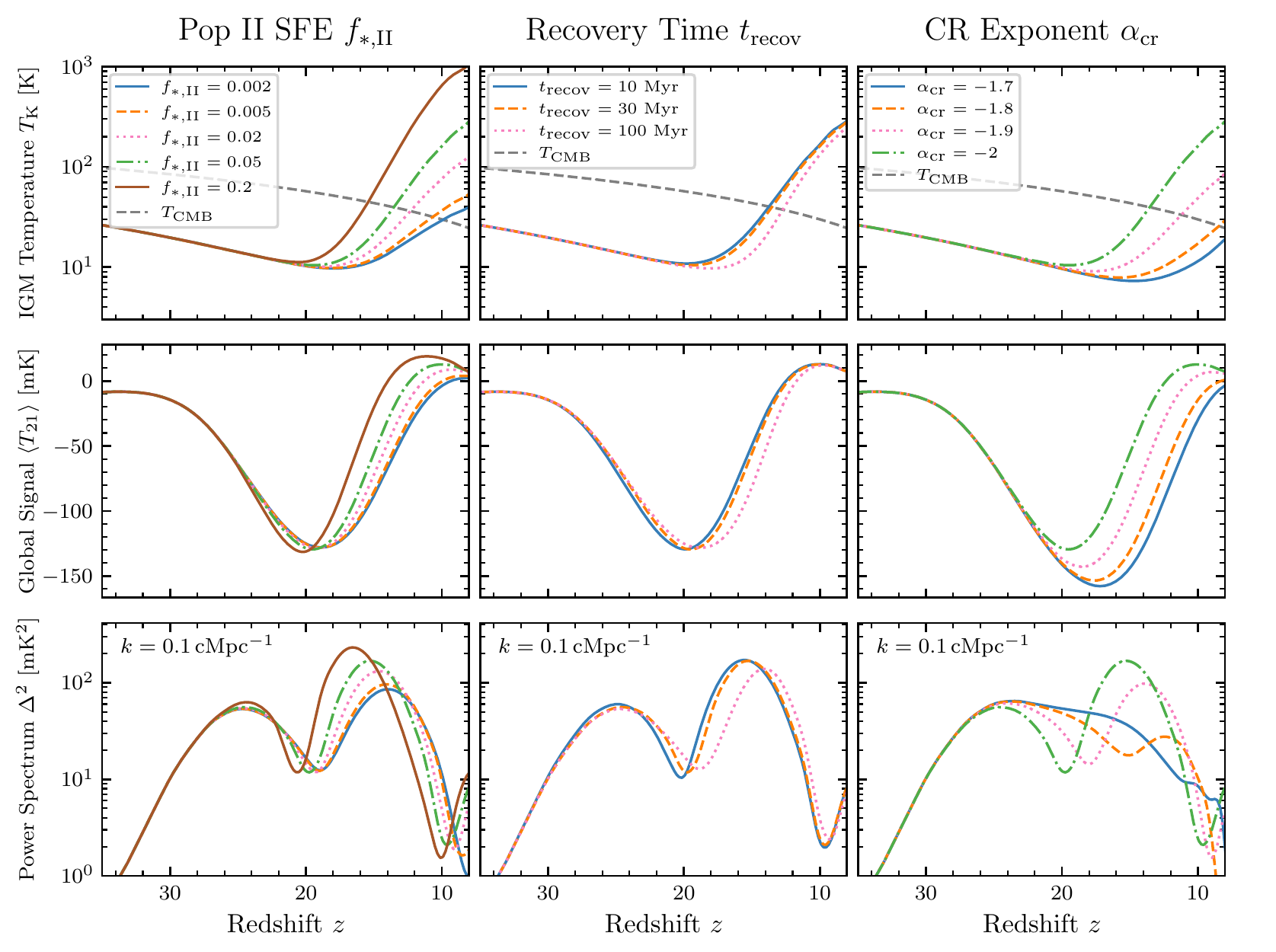}
    \caption{Variation of the IGM temperature and the 21-cm signal with astrophysical and cosmic ray parameters for diffusive escape cosmic rays. The kinetic temperature of the IGM (top row), the global 21-cm signal (middle row), and the 21-cm power spectrum at $k = 0.1$\,cMpc$^{-1}$ (bottom row) are shown for varying Pop II star formation efficiency $f_{\rm *, II}$ (left column), recovery time between Pop III and Pop II star formation $t_{\rm recov}$ (middle column), and exponent of the cosmic ray spectrum  $\alpha_{\rm cr}$ (right column). $f_{\rm *, II}$ is found to have a strong impact on the rate of the IGM heating and thus the 21-cm signal. We find an increase in the star formation efficiency from 0.002 to 0.2 results in the IGM temperature at $z = 8$ raising from 39.7 to 1070\,K. Conversely, $t_{\rm recov}$ is found to only have a small impact. Larger values of $t_{\rm recov}$ delay Pop II star formation and thus the IGM heating, leading to $\Delta z \sim 1$ shifts in the 21-cm global signal and power spectrum across the range of $t_{\rm recov}$ considered. The depicted variations with $\alpha_{\rm cr}$ replicate the findings of~\citet{Leite_2017} that the cosmic ray heating rate is quite sensitive to the spectral shape of the cosmic rays injected into the IGM, with increasing $\alpha_{\rm cr}$ from $-2$ to $-1.7$ reducing the $z = 8$ IGM temperature from 284 to 19\,K, and annihilating the power spectrum heating peak.}
    \label{fig:diffusive_escape_parameter_variation}
\end{figure*}

We start by considering models of cosmic ray heating similar to that proposed by \citet{Leite_2017}, with cosmic rays emitted by all stellar populations from all star-forming halos assuming $\eta_{\rm cr} = 10^{48}$\,erg\,\isolm{}, $\kemin{} = 10^{-2}$\,MeV, $\kemax{} = 10^9$\,MeV and $\alpha_{\rm cr} = -2$. As all stellar populations are emitting cosmic rays in this model, we anticipate heating to be dominated by Pop II stars given their greater star formation rate at later times. We thus consider the variation of the 21-cm signal with the parameters that determine the efficiency and timing of Pop II star formation, $f_{\rm *, II}$ and $t_{\rm recov}$ respectively. In addition, we seek to verify the finding of \citet{Leite_2017} that cosmic ray heating is quite sensitive to the spectral exponent of the cosmic ray spectrum by varying $\alpha_{\rm cr}$ as well. To isolate the impacts of the variation in each one of the aforementioned parameters we show $T_{\rm K}$ and the 21-cm signal as separate columns in Fig.~\ref{fig:diffusive_escape_parameter_variation} for different values of $f_{\rm *,II}$, $t_{\rm recov}$ and $\alpha_{\rm cr}$.

As expected, increasing the Pop II star formation efficiency increases the rate of cosmic ray heating in our simulations. Thus as $f_{\rm *,II}$ rises from $0.002$ to $0.2$ we find the redshift of $T_{\rm K}-T_{\rm cmb}$ equality moves to higher redshifts, from $z = 9.9$ to $z = 15.4$, and the temperature of the IGM at $z = 8$ increases from $40$ to $1070$\,K. This faster heating of the IGM is reflected in the 21-cm signal with higher $f_{\rm *,II}$ leading to a shallower absorption trough shifted to higher redshifts and a stronger emission peak. Similarly, as $f_{\rm *,II}$ increases the power spectrum heating peak shifts to higher redshifts and is stronger.

The effect of $t_{\rm recov}$ is found to be more modest. Higher values of the recovery time delay the onset of IGM heating by $\Delta z \sim 1$. As a result, for the range of $t_{\rm recov}$ considered here, the IGM always reaches a similar temperature of $257$ to $291$\,K at $z = 8$. These small shifts in the IGM thermal history lead to small changes in the 21-cm global signal and power spectrum with $t_{\rm recov} = 100$\,Myr having a global signal minimum and power spectrum heating maximum at corresponding delays of $\Delta z = 1$ and $\Delta z = 2$ compared to the equivalent features for $t_{\rm recov} = 10$ and $30$\,Myr. The discrepancies in the 21-cm signals for the explored $t_{\rm recov}$ values are similar to, if smaller in magnitude than, the differences seen by \citet{Magg_2022} when they considered the same parameter variation assuming X-ray heated models. The smaller differences observed in our work can be attributed to the cosmic ray heating being less efficient than the X-ray heating in \citet{Magg_2022}. Additional effect comes from our updated LW feedback prescription discussed earlier (see subsection \ref{subsec:LW}). 

Our simulations are found to support the finding of \citet{Leite_2017} that cosmic ray heating is quite sensitive to the spectral exponent $\alpha_{\rm cr}$ of the cosmic rays injected into the IGM. The aforementioned $T_{\rm K} > 250$\,K at $z = 8$ is achieved for a spectrum flat in $\log(\ke{})$ with $\alpha_{\rm cr} = -2$. As $\alpha_{\rm cr}$ increases a greater portion of the cosmic ray energy is in the form of higher energy cosmic rays ($\ke{} > 200$\,MeV), which as we found previously do not efficiently transfer their energy to the neutral IGM. Consequently, the efficiency of cosmic ray heating decreases quite sharply as $\alpha_{\rm cr}$ increases, with $T_{\rm K}$ only exceeding $T_{\rm CMB}$ at $z = 7.4$ in the case of an exponent with $\alpha_{\rm cr}  = -1.7$. The decrease in heating efficiency with $\alpha_{\rm cr}$ leads to deeper absorption troughs shifted to lower redshifts and the gradual elimination of the power spectrum heating peak. 

The strong dependence of the efficiency of cosmic ray heating on the spectrum of the injected cosmic rays suggests that accurate modeling of the escape mechanism is required to understand the role of cosmic rays in thermal history. Such simulations are made more challenging by the uncertainty in the strength of magnetic fields in the early universe and the nature of the astrophysical population in the first star-forming halos~\citep{Jana_2018}. Detection of cosmic ray heating through 21-cm observations would thus potentially provide insight into the magnetic field environment within the star-forming regions prior to reionization, as we would be able to test whether or not the low-energy cosmic rays were able to escape from their parent halos.

\subsubsection{Direct injection}~\label{ssec:direct}

Until now we have assumed cosmic rays can escape from halos of any mass 
following the diffusive escape model discussed in subsection~\ref{ssec:cr_escape}. However, \citet{Sazonov_2015} proposed an intriguing alternative wherein cosmic rays are released directly into the IGM due to the energetic SNRs of Pop III stars escaping their fully photoevaporated host halos. Since in this mechanism the SNR has to escape the host halo for cosmic rays to be directly injected into the IGM, this sets an upper limit on the mass of star-forming halos that contribute to cosmic ray heating, which in their work \citet{Sazonov_2015} take as $10^7$\,\solm. 

\begin{figure*}
	\includegraphics[width=\textwidth]{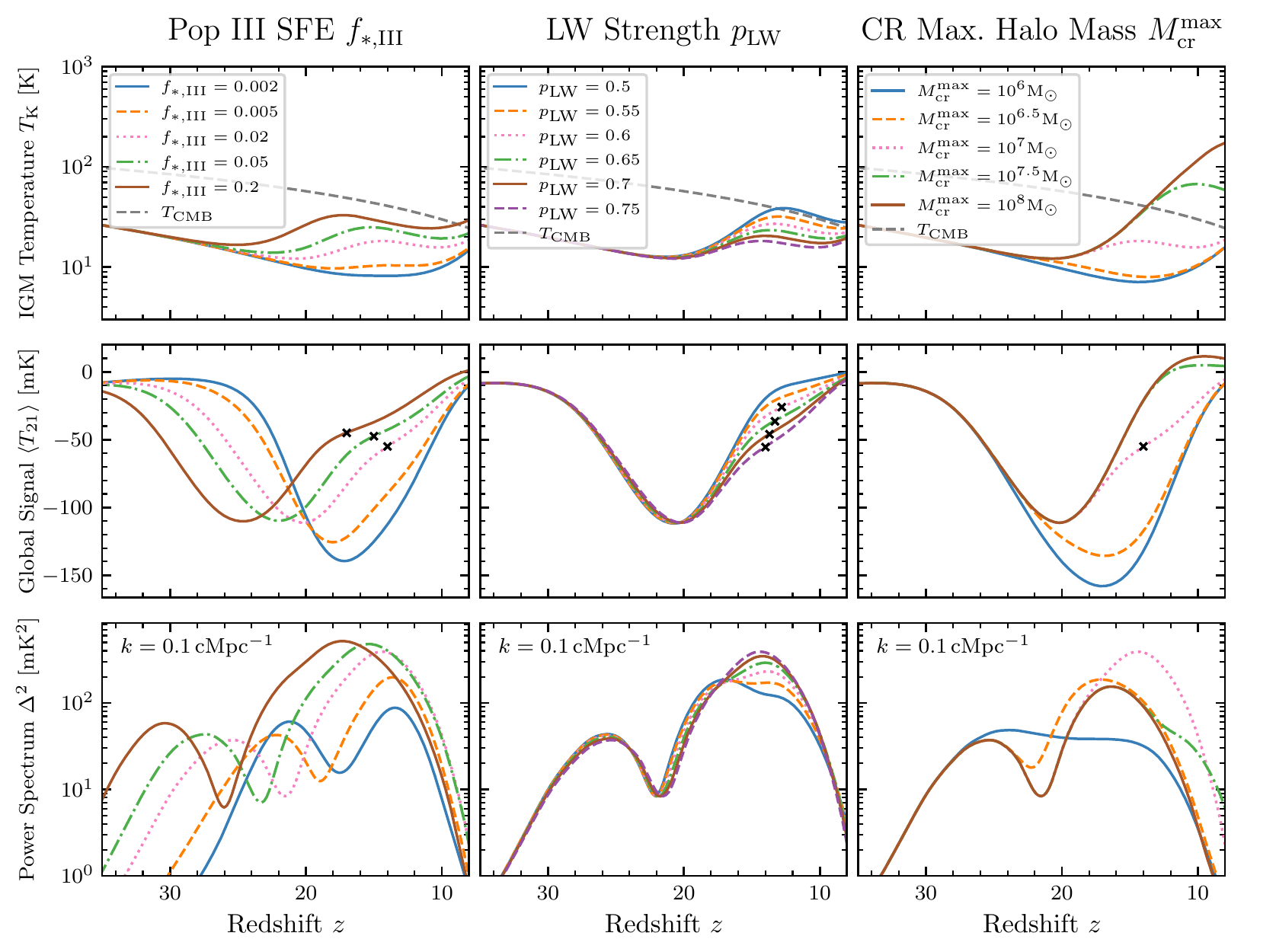}
    \caption{Variation of the IGM temperature and the 21-cm signal with astrophysical and cosmic ray parameters for the direct injection cosmic rays model. The kinetic temperature of the IGM (top row), the global 21-cm signal (middle row), and the 21-cm power spectrum at $k = 0.1$\,cMpc$^{-1}$ (bottom row) are shown for varying Pop III star formation efficiency $f_{\rm *,III}$ (left column), LW feedback delay parameter $p_{\rm LW}$ (middle column), and maximum halo mass cosmic rays can escape from $M_{\rm cr}^{\rm max}$ (right column). Some of the models are found to display a complex thermal history, with a maximum and a secondary minimum in $T_{\rm K}$. The height and redshift of this $T_{\rm K}$ maximum are found to be quite sensitive to $p_{\rm LW}$ and $M_{\rm cr}^{\rm max}$. This is due to the maximum being caused by  $M_{\rm mol, crit}$  (the critical halo mass for molecular cooling star formation) being increased above $M_{\rm cr}^{\rm max}$ by the LW feedback. Due to the LW feedback cutting off cosmic ray heating, we do not find it to be as efficient as was found in the previous study \citep{Sazonov_2015}. These complex IGM thermal histories in turn produce unusual inflexion points in the heating arm of the 21-cm signal (marked with black crosses), which we explore further in Fig.~\ref{fig:flattened_profile}.}
    \label{fig:direct_emission_parameter_variation}
\end{figure*}

To allow for easier comparison between our findings and those of \citet{Sazonov_2015} in this subsection we assume cosmic rays can only be injected into the IGM by Pop III stars, turn off X-ray heating, and adopt the same values for $\kemin{} = 10^{-3}$\,MeV, $\kemax{} = 10^8$\,MeV, and $\alpha_{\rm cr} = -2$ as in their work. As reference values we take $M_{\rm cr}^{\rm max} = 10^7$\,\solm, to match the aforementioned study, and use $\eta_{\rm cr} = 3 \times 10^{48}$\,erg\,\isolm{}, which is in the middle of the range considered by \citet{Sazonov_2015}. Since in this model cosmic rays are sourced from Pop III star supernovae, we consider the variation in cosmic ray heating with Pop III star formation efficiency $f_{\rm *,III}$ and the strength of the LW feedback \citep[set by the delay parameter $p_{\rm LW}$, see][]{Fialkov_2013} due to the LW feedback impacting the minimum halo mass in which Pop III stars can form. Finally, we also consider the sensitivity of this model to its unique feature, the halo mass threshold below which cosmic rays can escape into the IGM $M_{\rm cr}^{\rm max}$. Similarly to the last section, we calculate the IGM kinetic temperature, global 21-cm signal, and power spectrum for variations of these three parameters and show the results in Fig.~\ref{fig:direct_emission_parameter_variation}. 

For many of the parameter values we consider, including our fiducial values, an unexpectedly complex redshift evolution of the IGM kinetic temperature occurs. Rather than the familiar picture in which the gas temperature is driven by monotonic heating preceded by adiabatic cooling  (as we found in the previous subsection Fig.~\ref{fig:diffusive_escape_parameter_variation}), the direct injection models develop a local peak in the gas temperature at $z\sim 10-20$. In scenarios with the aforementioned complex $T_{\rm K}$ evolution, we find adiabatic cooling initially dominates the IGM temperature evolution (as expected). Around the first minimum in gas temperature ($z \gtrsim 20$) cosmic ray heating becomes efficient causing the kinetic temperature to rise. However, soon after cosmic ray heating becomes efficient, $M_{\rm mol, crit}$ rises above $M_{\rm cr}^{\rm max}$ due to the LW feedback (which was not modelled in \citet{Sazonov_2015}, hence they did not observe this unusual behaviour). Consequently, cosmic rays can no longer escape from star-forming halos, and so cosmic ray heating of the IGM is cut off. Adiabatic cooling thus begins to dominate the IGM temperature evolution for the second time in cosmic history,  the IGM temperature drops creating a local maximum in the thermal history. Finally, \lya{} heating becomes efficient at later times causing the secondary minimum in $T_{\rm K}$ as the IGM temperature rises. \lya{} heating remains efficient for the rest of our simulation resulting in a further monotonically increasing gas temperature. 

In Fig.~\ref{fig:direct_emission_parameter_variation} we explore how this peculiar behaviour changes with the efficiency of the LW feedback. The differences primarily come from the fact that  $p_{\rm LW}$ regulates the redshift at which $M_{\rm mol, crit}$ exceeds the maximal halo mass for cosmic ray escape. As we increase $p_{\rm LW}$ the LW feedback is stronger and so  $M_{\rm mol, crit}$ exceeds $M_{\rm cr}^{\rm max}$ earlier. Consequently, the peak in gas temperature is shifted to earlier times and is lower due to the shorter phase of cosmic ray heating. Curiously, the presence of the peak in $T_{\rm K}$ is manifested in the global 21-cm signal as an additional inflexion point (marked by black crosses in Fig.~\ref{fig:direct_emission_parameter_variation}). We find that an earlier and weaker peak in $T_{\rm K}$ in simulations with larger $p_{\rm LW}$ values leads to the inflexion point moving to higher redshifts. In addition, models with stronger LW feedback (larger $p_{\rm LW}$) produce  deeper  global 21-cm signals due to less efficient cosmic ray heating. This effect is also manifested in the power spectra as enhanced power at and around the heating peak. 

Next, we explore how changing $M_{\rm cr}^{\rm max}$ affects the shape of the signals. Decreasing the free parameter $M_{\rm cr}^{\rm max}$ increases the redshift at which $M_{\rm mol, crit}  = M_{\rm cr}^{\rm max}$ thus leading to an earlier cut-off of cosmic ray heating. For the lowest explored value of $M_{\rm cr}^{\rm max} \sim 10^6$\,\solm{}, cosmic ray heating is cut-off very early (at $z > 20$) and before its rate can exceed the adiabatic cooling rate.  As a result, in this model, the IGM continues to cool down adiabatically until \lya{} heating becomes efficient at $z \sim 13$. The resulting thermal history is standard (without visible local maximum/minima). This picture qualitatively changes as $M_{\rm cr}^{\rm max}$ rises above $\sim 10^7$\,\solm{}. In such scenarios, cosmic ray heating is efficient for a sufficient amount of time to imprint changes in thermal history. At extremely high values of $M_{\rm mol, crit}$ (at and above $\sim 10^8$\,\solm{}) cosmic ray heating becomes the dominant mechanism and is not cut off even by $z = 8$. In this simulation the resulting IGM temperature is monotonically rising after the onset of cosmic ray heating without developing any unexpected features. The resulting 21-cm global signal and power spectrum are also quite sensitive to the choice of $M_{\rm cr}^{\rm max}$ with low values ($M_{\rm cr}^{\rm max} = 10^6$\,\solm{}) resulting in a deep and late absorption trough ($-158$\,mK at $z = 17$). Owing to the absence of efficient cosmic ray heating in this case the heating peak in the power spectrum is very weak. For the intermediate values  ($M_{\rm cr}^{\rm max} \sim 10^7$\,\solm{}) we find a more complex behaviour with additional inflexion points in the global 21-cm signal between $z = 20$ and $10$ and an enhanced power spectrum peak. Finally, as expected from the thermal history, the highest explored values ($10^8$\,\solm{}) with efficient heating result in an earlier and shallower absorption trough ($-111$\,mK at $z = 20$) and a well-defined power spectrum heating peak. 

The impact of $f_{\rm *,III}$ on the 21-cm signal is more complex than for the other two explored parameters since it affects both the rate of cosmic ray heating and the strength of the LW background. In our simulations, we find the local peak in $T_{\rm K}$ occurring for models with $f_{\rm *,III} \gtrsim 0.02$, while models with lower values of $f_{\rm *,III}$ feature a flattened extended $T_{\rm K}$ minimum. This dependence illustrates that the stronger LW feedback, and thus the earlier cut-off of cosmic ray heating (e.g., from more vigorous Pop III star formation), is not sufficient to fully counteract the more efficient cosmic ray heating resulting from the higher Pop III star formation efficiency. As a result, we observe an earlier but higher $T_{\rm K}$ peak as $f_{\rm *,III}$ increases (in the regime $f_{\rm *,III} \gtrsim 0.02$). Consequently, higher $f_{\rm *,III}$ values lead to earlier and slightly shallower absorption troughs with inflexion points at progressively higher redshifts owing to the increasingly more efficient cosmic ray heating. On the other hand, we find that inefficient Pop III formation with the values  $f_{\rm *,III} < 0.005$ results in a deep global 21-cm signal with no additional inflexion point. The power spectrum trends are somewhat simpler, with higher $f_{\rm *,III}$ leading to earlier and stronger peaks from both the increased \lya{} emission (peak at $z > 20$) and the more efficient heating (peak at $12 < z < 17$). We observe no apparent new behaviour in the power spectra, unlike in $T_{\rm K}$ and $\langle T_{\rm 21} \rangle$,  around $f_{\rm *,III} \sim 0.02$.

We find that indeed direct injection of cosmic rays can heat up the IGM. However, we do not find this particular model of cosmic ray heating to be as efficient as in the original study by \citet{Sazonov_2015}. In part, this can be attributed to differences in star formation prescription, the relaxation of the global heating assumption that we introduced here, and the truncation of cosmic ray heating by the LW feedback. Consequently, whereas \citet{Sazonov_2015} found cosmic rays to heat the IGM by $10$ to $100$\,K at $z = 15$, in our simulations $10$\,K of heating at $z = 15$ is only achieved for high values of star formation rates ($f_{\rm *, III} \geq 0.02$) or high escape mass thresholds ($M_{\rm cr}^{\rm max} \geq 10^7$\,\solm{}) with the largest $z = 15$ temperature increase above the adiabatic cooling solution $\Delta T_{\rm K} = 20.3$\,K achieved for $M_{\rm cr}^{\rm max} = 10^8$\,\solm{}.
While the amount of heat deposited into the IGM by cosmic rays is found to be smaller than in \citet{Sazonov_2015}, the 21-cm signal is still sensitive to this mechanism. Our results suggest that the 21-cm signal could provide a probe of the nature of Pop III star supernovae through the small amount of cosmic ray heating which they produce around $z = 15-20$.

\begin{figure*}
	\includegraphics[width=\textwidth]{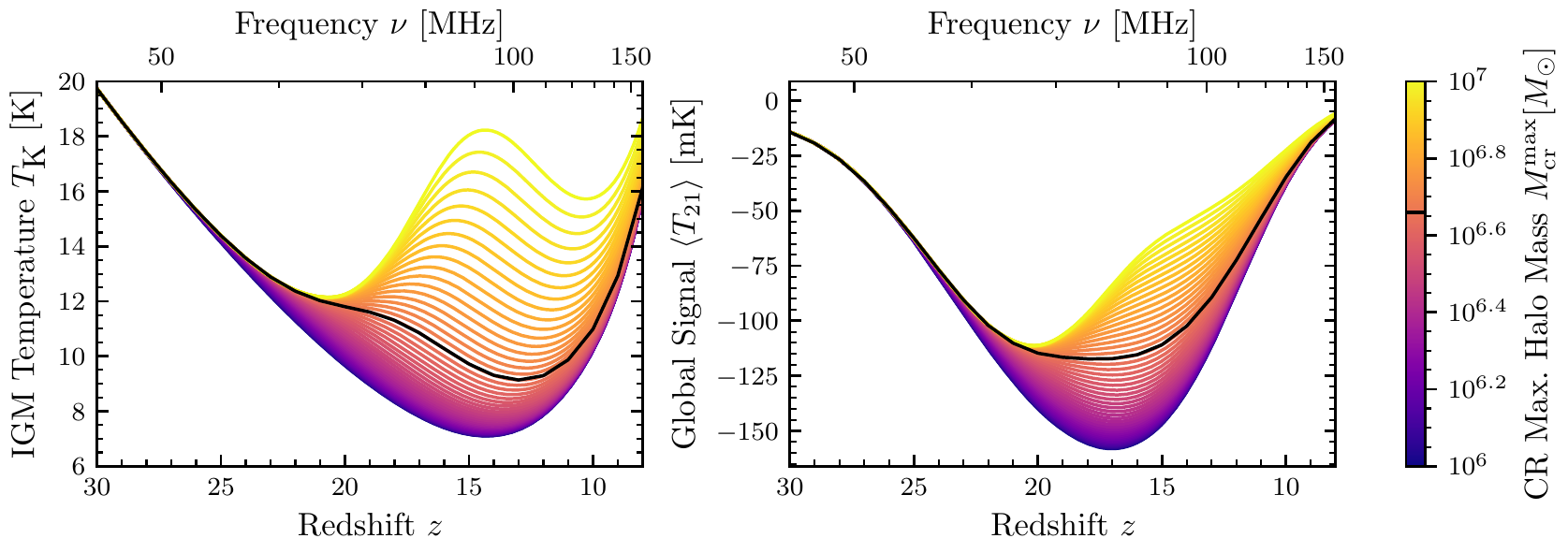}
    \caption{Flattened global 21-cm signal minimum from the cosmic ray heating cut off. Variation of the IGM kinetic temperature and the global 21-cm signal is shown colour-coded with respect to the maximum halo mass that cosmic rays can escape from. At the highest end of the range considered, $M_{\rm cr}^{\rm max} =10^7$\,\solm, $T_{\rm K}$ develops a local peak due to the contribution of cosmic rays to heating. In turn, an inflexion point appears in the global signal. As $M_{\rm cr}^{\rm max}$ is decreased the $T_{\rm K}$ maximum merges with the first minimum, and the inflexion point in the global signal coincides with the minimum of the absorption trough producing a flattened absorption feature at $M_{\rm cr}^{\rm max} = 4.57 \times 10^6$\,\solm{}(highlighted in black). For lower values of the critical mass, the global signal minimum gets narrower and deeper and is no longer flattened as cosmic ray heating becomes insignificant.}
    \label{fig:flattened_profile}
\end{figure*}

To recap, we find that the direct injection model can lead to a rich IGM thermal history and unusual features in the 21-cm signal. Above we showed that the shape of the global signal undergoes a qualitative transition as the maximal halo mass changes from  $M_{\rm cr}^{\rm max} = 10^6$ to $10^7$\,\solm{}. A new inflexion point is developed as a manifestation of the complex heating history owing to the contribution of cosmic rays. Now we consider this transition in more detail by densely sampling the parameter  $M_{\rm cr}^{\rm max}$. We run the code 50 times with equal logarithmic spacing in $M_{\rm cr}^{\rm max}$ in the range $10^6-10^7$. The results for $T_{\rm K}$ and $\langle T_{\rm 21} \rangle$  are shown in Fig.~\ref{fig:flattened_profile}.

Examining the resulting thermal histories we find a clear peak in temperature for $M_{\rm cr}^{\rm max} = 10^7$\,\solm{}, the upper end of our considered range. As $M_{\rm cr}^{\rm max}$ decreases, the peak shifts to higher redshifts and lower temperature values. Eventually, the local maximum merges with the higher redshift minimum transforming into two inflexion points that can be seen in the highlighted signal (black line) corresponding to $M_{\rm cr}^{\rm max} = 4.57 \times 10^6$\,\solm. For lower  values of the critical mass, thermal evolution shows the classical behaviour driven by the adiabatic cooling and \lya{} heating, with no extra features and no clear sign of cosmic ray heating. 

Considering the effect on the global 21-cm signal, we see that as $M_{\rm cr}^{\rm max}$ decreases the inflexion point shifts to higher redshifts. Eventually, the inflexion point merges with the minimum of the absorption trough producing a flattened profile (black line in the right panel of Fig.~\ref{fig:flattened_profile}). As we reduce the value of $M_{\rm cr}^{\rm max}$ even further, the absorption trough becomes deeper, narrower, and is no longer flattened.  

Intriguingly, the flattened absorption feature that we find for  $M_{\rm cr}^{\rm max} = 4.57 \times 10^6$\,\solm{} has a similar shape to the disputed EDGES Low Band detection of the global signal~\citep{EDGES} which has a flattened minimum at a similar redshift. We note, however, that cosmic ray heating cannot provide a self-consistent explanation of the EDGES best-fit signal on its own as the resulting absorption trough does not have the required depth
or the steep sides observed by EDGES. However, it does illustrate that such flattened global signals are not inherently pathological and can be achieved in models with two separate heating mechanisms, one active at higher redshifts and only mildly pre-heating the IGM and the other representing a more sustained heating that becomes efficient at later times (such as X-ray heating, \lya{} or CMB heating). The high-redshift preheating could be achieved also in other models such as e.g., fractional annihilating dark matter with a cross-section that decreases with time~\citep{Liu_2018}.

\section{Discussion and Conclusions}
\label{sec:conclusions}

In this paper, we have developed a model of cosmic ray heating for use in semi-numerical 21-cm signal simulations.
Unlike previous globally-averaged models our approach includes realistic modelling of the spatial distribution of cosmic ray heating.
We bracket the range of possible cosmic ray heating scenarios by considering free-streaming versus locally-confined models of particle propagation, exploring the dependence of heating on the cosmic ray energy spectrum, and considering different channels for escape from host halos. 
Our method thus allows us to perform the first investigations of the signatures of this heating mechanism in 21-cm tomographic maps and power spectra, as well as the previously studied global 21-cm signal. 
It also enables us to compare the contribution of cosmic rays to the thermal history of the IGM to other heating mechanisms, such as X-ray heating, and thus to propose novel methods to potentially distinguish an X-ray heated IGM from a cosmic-ray heated IGM using upcoming observations.

We find the spatial range of cosmic ray heating to be much shorter than that of X-ray heating, even when cosmic rays are assumed to travel along straight lines. This short-range nature manifests in the IGM heating being clustered around regions of efficient star formation and results in a sharp contrast in the 21-cm signal between localized heated regions of emission and vast unheated regions of absorption. 

Comparing our inhomogeneous cosmic ray heating model to a uniform model commonly used in the literature, we find that its localized nature results in a slightly delayed evolution of the global 21-cm signal. This demonstrates the existence of a small assumption-induced bias in the results of previous globally-averaged studies.
Much larger differences are seen in the power spectra and tomographic maps, as with cosmic rays as a dominant heating source the 21-cm signal traces matter fluctuations to smaller scales than in the case of X-rays. 
Specifically, we find that the 21-cm power spectrum at higher $k$ is increasingly suppressed for the longer-ranged heating mechanisms allowing us to constrain the propagation mechanism of cosmic ray particles and distinguish cosmic ray from X-ray heating. 
Any clustering of cosmic ray heating on scales smaller than our simulation resolution would enhance these effects and thus strengthen our conclusions.
The potential of the high-$k$ end of the power spectrum to provide information about the spatial range of the dominant heating mechanism is anticipated to apply more generally and be robust to astrophysical uncertainties, making it a useful diagnostic tool. For example, it has previously been suggested~\citep{Fialkov_2014c, Fialkov_2014b} that the high-$k$ end of the power spectrum could be used to probe the SED of early X-ray binaries.

Due to the uncertainties surrounding cosmic ray escape into the IGM, we explored several plausible mechanisms. In models where cosmic rays can escape from halos of any mass via diffusion or outflow advection, we found cosmic ray heating of the IGM can be more efficient than X-ray heating. However, similarly to \citet{Leite_2017}, we found that the efficiency of cosmic rays as a heating source is strongly dependent on the spectrum of the particles injected into the IGM. A change in the source kinetic energy spectral exponent from $-2$ to $-1.7$ decreases the IGM temperature from $284$ to $19$\,K at $z = 8$. This is because only the lower energy cosmic ray protons are able to efficiently transfer their kinetic energy to the IGM as heat, as was previously demonstrated in \citet{Sazonov_2015}.

For cosmic ray heating models similar to that of \citet{Sazonov_2015}, wherein cosmic rays can only escape from lower mass Pop III star-forming halos, we found cosmic ray heating to be relatively inefficient. 
Due to \citet{Sazonov_2015} not considering the LW feedback in their model, we were only able to replicate the significant increase in temperature they observed for unphysically weak LW feedback or if we raised the threshold halo mass of cosmic ray sources. 
Instead at theoretically motivated parameter values we find the interplay between this mass threshold and the LW feedback, which raises  $M_{\rm mol, crit}$, leads to unexpectedly complex thermal histories. In such models, the IGM kinetic temperature was found to have a local peak due to cosmic ray heating being cut off by the LW feedback leading to a secondary episode of cosmic cooling before \lya{} heating from Pop II stars becomes efficient and re-heats the IGM. This unusual thermal evolution results in an inflexion point in the heating arm of the global 21-cm  signal. By fine-tuning the mass threshold for cosmic ray escape, this inflexion point can be merged with the minimum of the absorption trough resulting in a flattened profile. Thus, we demonstrate that flat-based global 21-cm signals (like the one detected by EDGES) can in principle be achieved in models where one heating mechanism becomes inefficient at a similar time to the gas becoming fully coupled via the WF effect and another heating mechanism turns on at a later time. We do not advocate for this model to be a self-consistent explanation of the  EDGES best-fit signal as it does not explain the depth or the steep sides of the detected absorption trough (although a dual-heating scenario could form a part of a more complex explanation which also invokes a radio contribution to create the deep absorption). 

There remains a major uncertainty in cosmic ray heating, namely the strength of the IGM primordial magnetic field which scatters cosmic rays. This uncertainty leads to a large uncertainty in the cosmic ray diffusivity within the IGM. Furthermore, in the case of primordial magnetic fields at the higher end of the experimentally allowed range ($> 0.01$\,nG comoving field strength), the transfer of energy from cosmic rays to the IGM via Alfv\'en waves~\citep{Bera_2022} and synchrotron radio emission from cosmic ray electrons~\citep{Jana_2019} are anticipated to become efficient. Since these effects are only efficient for the largest experimentally allowed primordial magnetic fields, we do not include them in our modelling. However, should the primordial magnetic field take such a high value, these effects would result in stronger heating of the IGM from cosmic rays, clustered around star-forming halos, and enhanced radio background, the former acting to diminish the 21-cm signal and the latter enhancing it. This sensitivity of the 21-cm signal, in particular the 21-cm power spectrum, to cosmic ray heating of the IGM provides an indirect probe of the primordial IGM magnetic field.

Overall we found that the distinct nature of cosmic ray heating means it leaves characteristic signatures in 21-cm tomographic maps, the high-$k$ end of the power spectrum, and in some cases the 21-cm global signal. These features are of sufficient magnitude to potentially be probed by the 21-cm signal measurements from current and next-generation experiments. Thus, an understanding of cosmic ray heating is necessary for the correct interpretation of the 21-cm signal, and conversely, we may be able to constrain early universe astrophysics that impacts cosmic rays (for example, the primordial IGM magnetic field) through the 21-cm signal.

\section*{Acknowledgements}

The authors would like to thank Donghui Huang for helpful conversations concerning the updating of the LW feedback prescription in our simulation code, as well as to express our gratitude to John Cumner and Boyuan Liu for their invaluable comments on earlier versions of this paper. 
We would additionally like to thank the anonymous referee, for their careful reading of this work and insightful comments, that have led to a much improved final manuscript.

TGJ would like to thank the Science and Technology Facilities Council (UK) for their continued support through grant number ST/V506606/1. 
AF is supported by a Royal Society University Research Fellowship \#180523.
EdLA acknowledges the support of the Science and Technology Facilities Council (UK) through a Rutherford Fellowship.
WJH thanks the Royal Society for their support through a Royal Society University Research Fellowship. 
RB acknowledges the support of the Israel Science Foundation (grant No.\ 2359/20), the Ambrose Monell Foundation, the Institute for Advanced Study, the Vera Rubin Presidential Chair in Astronomy, and the Packard Foundation.

For the purpose of open access, the author has applied a Creative Commons Attribution (CC BY) licence to any Author Accepted Manuscript version arising from this submission.

\section*{Data Availability}

The code to calculate the window functions used in this study to model cosmic ray propagation is made available at \href{https://github.com/ThomasGesseyJones/CosmicRayHeatingFor21cm}{https://github.com/ThomasGesseyJones/CosmicRayHeatingFor21cm}.
All other data used and generated in the writing of this article will be shared on reasonable request to the corresponding author.



\bibliographystyle{mnras}
\bibliography{cosmic_rays} 








\bsp	
\label{lastpage}
\end{document}